\documentclass[aoas]{imsart}

\usepackage{amsmath}
\RequirePackage{amsthm,amsmath,amsfonts,amssymb}
\RequirePackage[authoryear]{natbib}
\RequirePackage[hidelinks]{hyperref}
\RequirePackage{graphicx}

\startlocaldefs
\theoremstyle{plain}

\newtheorem{theorem}{Theorem}[section]

\newtheorem{corollary}[theorem]{Corollary}
\theoremstyle{remark}

\usepackage{bbm}
\usepackage{mathtools}
\usepackage{multirow}
\usepackage{multicol}
\usepackage[section]{algorithm}
\usepackage{algorithmicx}
\usepackage{algpseudocode}
\usepackage{algcompatible}
\usepackage{stackrel}
\usepackage[flushleft]{threeparttable}
\usepackage{tikz}
\usetikzlibrary{shapes,arrows,decorations.markings}
\usetikzlibrary{decorations.pathreplacing,calc}
\usetikzlibrary{arrows.meta}
\usetikzlibrary{positioning}
\usepackage{mathtools}
\usepackage[many]{tcolorbox}
\newtcolorbox{rightbrace}{
	enhanced jigsaw, 
	breakable, 
	frame hidden,
	parbox=false,
}

\algblock{Input}{EndInput}
\algnotext{EndInput}
\algblock{Output}{EndOutput}
\algnotext{EndOutput}
\def\phaseone{phase-I}
\def\phasetwo{phase-II}
\newcommand{\ind}{\perp\!\!\!\perp}
\endlocaldefs

\begin{document}

\begin{frontmatter}
\title{
Patient recruitment using electronic health records under selection bias: a two-phase sampling framework}
\runtitle{Patient Recruitment Using EHR}

\begin{aug}
\author[A]{\fnms{Guanghao} \snm{Zhang}\ead[label=e1,mark]{ghzhang@umich.edu}},
\author[B]{\fnms{Lauren J.} \snm{Beesley}\ead[label=e2]{lvandervort@lanl.gov}},
\author[A]{\fnms{Bhramar} \snm{Mukherjee}\ead[label=e3,mark]{bhramar@umich.edu}},
\and
\author[A]{\fnms{Xu} \snm{Shi}\ead[label=e4,mark]{shixu@umich.edu}}
\address[A]{Department of Biostatistics, University of Michigan,
\printead{e1,e3,e4}}

\address[B]{Statistical Sciences Group,
Los Alamos National Laboratory,
\printead{e2}}
\end{aug}

\begin{abstract}
Electronic health records (EHRs) are increasingly recognized as a cost-effective resource for patient recruitment in clinical research. However, how to optimally select a cohort from millions of individuals to answer a scientific question of interest remains unclear. Consider a study to estimate the mean or mean difference of an expensive outcome. Inexpensive auxiliary covariates predictive of the outcome may often be available in patients' health records, presenting an opportunity to recruit patients selectively which may improve efficiency in  downstream analyses. In this paper, we propose a two-phase sampling design that leverages available information on auxiliary covariates in EHR data. A key challenge in using EHR data for multi-phase sampling is the potential selection bias, because EHR data are not necessarily representative of the target population. Extending existing literature on two-phase sampling design, we derive an optimal two-phase sampling method that improves efficiency over random sampling while accounting for the potential selection bias in EHR data. We demonstrate the efficiency gain from our sampling design via simulation studies and an application to evaluating the prevalence of hypertension among US adults leveraging data from the Michigan Genomics Initiative, a longitudinal biorepository in Michigan Medicine.
\end{abstract}

\begin{keyword}
\kwd{auxiliary information}
\kwd{electronic health records}
\kwd{selection bias}
\kwd{study design}
\kwd{two-phase sampling}
\end{keyword}

\end{frontmatter}

\section{Introduction}\label{sec1}

Electronic health record (EHR) data are increasingly used to facilitate patient recruitment \citep{effoe2016use,cowie2017electronic,mc2019using}. 
	An EHR is a digital repository of routinely collected patient health information, including medical diagnosis, procedure, medication, radiology images, and laboratory test \citep{hayrinen2008definition,shortreed2019challenges}. 
	In conventional observational and randomized studies, patient recruitment and patient retention are often limited by funding and time. In addition, the recruited cohort tends to be homogeneous and not representative of a real-world population, thus study results are often not generalizable to a target population for health policy decision making \citep{hemkens2016routinely}.
	In contrast, the rich clinical information and real-world population provided by EHR present a cost-effective data source to identify and recruit patients who satisfy the eligibility criteria of a research study \citep{bower2017active,schreiweis2014comparison}.
	For example, \citet{wu2017semehr} developed a semantic search system that is used to recruit patients into the $100,000$ Genomes Project leveraging clinical notes in EHR; \citet{thadani2009electronic} deployed an electronic screening method to identify eligible patients for trial recruitment. 
However, patient recruitment using EHR data has been limited to random sampling after applying the inclusion/exclusion criteria. \textcolor{black}{Because EHR data from a healthcare system may be biased towards certain demographic groups or specific health conditions, random sampling may lead to a biased cohort. In addition, random sampling may not sufficiently capture a rare event and does not utilize information on risk factors recorded in EHR.} It remains unclear how to optimally select a cohort from millions of individuals to answer a scientific question of interest.
	
	The goal of this paper is to improve the usability of EHR data for patient recruitment and population-level parameter estimation with an efficient sampling design framework. 
	A motivating example is the Michigan Genomics Initiative (MGI), a longitudinal biorepository at the University of Michigan Health System that was launched in 2012, linking patient EHR data  with genetic data to facilitate biomedical research. Patients $18$ years of age or older who underwent surgery at the University of Michigan Health System are approached for enrollment. Participants provide broad opt-in consent for use of their EHR data and biospecimen, as well as recontact in the future for any applicable research study. As such, an increasing number of survey, experimental, and observational studies have been conducted by recruiting patients from the MGI cohort based on their medical records \citep{joyce2021associations,wu2021exposure}. For example, a sample of eligible MGI patients were surveyed in March and April 2020 to evaluate risk factors for COVID-19 and the impact of the `Stay Home Stay Safe' executive order on Michigan residents \citep{wu2021exposure}.
	
	With the growing availability of MGI participants, it is possible to selectively recruit eligible patients to obtain maximal information under a budget constraint. 
	We illustrate this by estimating the prevalence of a chronic disease, such as hypertension, in the US adult population using MGI data. 
	Typically outcome assessment is expensive or time-consuming, while inexpensive auxiliary covariates predictive of the outcome are readily available in EHR data. 
	Therefore, instead of randomly recruiting eligible patients, 
	investigators can leverage such auxiliary information to sample patients from MGI according to an optimized sampling probability to improve efficiency in downstream analyses. 

		Our proposal is motivated by the observation that patient recruitment using EHR data 
	constitutes a two-phase sampling framework.
	As illustrated in Figure~\ref{fig:appdata}, from a pre-specified target population, a subset of individuals seek healthcare and their information is recorded in 
	\begin{figure}[h]
    \centering
		\tikzstyle{block} = [rectangle, draw, fill=gray!20, 
		text width=10 cm, text centered, 
		rounded corners, minimum height=1cm]
		\tikzstyle{wideblock} = [rectangle, draw, fill=gray!20, text centered, 
		text width=15 cm, rounded corners, minimum height=1cm]
		\tikzstyle{midblock} = [rectangle, draw, fill=gray!20, text centered, 
		text width=8 cm, rounded corners, minimum height=1cm]
		\tikzstyle{line} = [draw, -latex']
		\tikzstyle{vecArrow} = [thick, decoration={markings,mark=at position
			1 with {\arrow[semithick]{triangle 60}}},
		double distance=1.4pt, shorten >= 5.5pt,
		preaction = {decorate},
		postaction = {draw,line width=1.4pt, white,shorten >= 4.5pt}]
		\tikzstyle{innerWhite} = [semithick, black,line width=1.4pt, shorten >= 4.5pt]
		\scalebox{0.9}{\begin{tikzpicture}[node distance = 3cm, auto]
							\node[block](a){\textbf{US adult population}\\(The target population) \\\text{Sample size: $n=330,000,000$}
							\begin{align*}
							&\text{Summary based on NHANES 2017-2018 data:}\\[-0.8em]
							&~~~~\text{$28.16\%$ $\geq$ $60$-years, $51.79\%$ female, $62.00\%$ white,}\\[-0.8em]
							&~~~~\text{$58.87\%$ never-smoker, $41.95\%$ obese}\\[-0.8em]
							&~~~~\text{Benchmark hypertension prevalence: $\boldsymbol{39.19\%}$}
                            \end{align*}\\
                            };
							\node[block, below=0.5cm of a, yshift=-0.5cm](b){\textbf{The EHR sample: MGI} \\{(Phase I)}\\\text{Sample size: $n_e=80,934$}
							\begin{align*}
							&\text{Summary based on MGI 2018 data:}\\[-0.8em]
							&~~~~\text{$47.51\%$ $\geq$ $60$-years, $53.92\%$ female, $84.49\%$ white,}\\[-0.8em]
							&~~~~\text{$54.72\%$ never-smoker, $42.93\%$ obese}\\[-0.8em]
							&~~~~\text{Hypertension prevalence in MGI: $\boldsymbol{49.19\%}$}
                            \end{align*}\\
                            };
					\node[block, below=0.5cm of b, yshift=-0.5cm](c){\textbf{The study sample} \\{(Phase II)}\\\text{Sample size: $n_s=100$ if $\text{budget}=10^5$} \\
					\qquad\qquad\qquad\: \text{$n_s=1,000$ if $\text{budget}=10^6$}
					\begin{align*}
					    &\text{Estimated hypertension prevalence and sample size:}\\[-0.8em]
					    &~~~~\text{$\boldsymbol{38.32\%}$ ($95\% \text{ CI: }28.90\%, 47.75\%$), if $\text{budget}=10^5$}\\[-0.8em]
					    &~~~~\text{$\boldsymbol{39.47\%}$ ($95\% \text{ CI: }36.61\%, 42.32\%$), if $\text{budget}=10^6$}
					\end{align*}\\};

							\path [line] (a) -- (b)-- node[midway, right]
							{Address {\phaseone} selection bias (Section \ref{subsec:bias})}(a);
							\path [line] (b) -- (c)-- node[midway, right]  
							{Optimal {\phasetwo} sampling (Section \ref{subsec:opt})}(b);
					
							\draw[vecArrow] (a) to (b);
					\draw[vecArrow] (b) to (c);
					
							\draw[innerWhite] (a) to (b);
					\draw[innerWhite] (b) to (c);
			\end{tikzpicture} }
	\caption{The two-phase sampling framework for selective patient recruitment using EHR
data from the MGI: illustrated by estimating hypertension prevalence in the US adult population. 
Hypertension is chosen as the outcome of interest because it is readily measured in MGI and NHANES data which allows for validation. 
The raw prevalence in MGI  (49.19\%) is much higher than the benchmark value computed based on NHANES (39.19\%), indicating selection bias in the EHR sample. 
We aim to optimally recruit patients from the MGI cohort into a study sample to improve efficiency in estimation of hypertension prevalence  while accounting for potential selection bias. 
\label{fig:appdata}}
\end{figure}
\noindent an EHR system. 
	We refer to such a cohort as the EHR sample or the {\phaseone} sample. 
	From the {\phaseone} EHR sample, we aim to recruit a subset of patients into the study sample, i.e., the {\phasetwo} sample in a way such that the ultimate results are generalizable to a target population of inference (e.g., the US adult population). A review of the two-phase sampling literature is provided in   Section 1 of the Supplementary Material \citep{zhang2022supp}.

	However, existing two-phase sampling methods cannot be directly applied because the {\phaseone} EHR sample such as the MGI cohort is not necessarily a random sample of the target population \citep{phelan2017illustrating,goldstein2016controlling,beesley2020emerging}.
	Patients are observed in EHR data only if they seek care. 
When the EHR sample differs from the target population in terms of  characteristics relevant to the scientific question in view, 
	parameter estimates could be biased and statistical conclusions may lack generalizability \citep{tripepi2010selection}. 
 Several methods to model and mitigate selection bias in EHR data have been developed recently. 
	For example, \citet{haneuse2016general} proposed to model the selection mechanism by breaking it down {into} submechanisms due to a sequence of decisions made by patients, providers, and healthcare systems. \citet{goldstein2016controlling} considered addressing selection bias in EHR by controlling for healthcare utilization measured by  number of medical encounters.  
	\citet{Beesley2019.12.26.19015859} proposed calibration weighting and inverse probability of selection weighting methods to account for the differences between patients included and not included in the EHR cohort. 

In this paper, we propose an optimal design for patient recruitment using EHR data under a two-phase sampling framework while accounting for potential selection bias in the EHR cohort.  
	We first present an estimator of the mean  outcome 
	that acknowledges selection bias and leverages auxiliary information in EHR data. 
	 We then derive the optimal sampling probability for recruiting patients from the EHR cohort into the study cohort which minimizes the variance of this estimator. 
	We extend existing two-phase sampling methods to further account for selection bias with two approaches: direct estimation of selection mechanism and indirect bias reduction via subsampling strategy. We ultimately provide a two-phase design framework for estimation of a general estimand that is the solution to a given estimating equation, such as the average treatment effect (ATE) when the {\phasetwo} study is either an observational study or a randomized controlled trial (RCT), and coefficients in a regression model. 
	
	The rest of the paper is organized as follows. Section~\ref{sec:prelim} formulates EHR-based patient recruitment into a two-phase sampling framework and presents design and estimation strategies under this framework. We detail our proposed methods in Section~\ref{sec:method}, where we derive the optimal {\phasetwo} sampling probability under a given budget allowing for a biased {\phaseone} EHR sample in Section~\ref{subsec:opt},  present two methods for addressing the selection bias in EHR data in  Section~\ref{subsec:bias}, and detail an estimator of the parameter of interest in Section~\ref{subsec:estbeta}. We prove the efficiency gain compared to random sampling in Section~\ref{sec:RE}. In Section~\ref{sec:ext}, we extend our proposed framework to facilitate estimation of a general estimand such as the ATE and regression coefficients.
		We demonstrate the efficiency gain via extensive simulation studies in Section~\ref{sec:simulation}, and we illustrate our proposed methods with an application to estimating the prevalence of hypertension in the US adult population using MGI data in Section~\ref{sec:application}. We conclude with a brief discussion in Section~\ref{sec:discussion}. 

\section{Preliminaries}
	\label{sec:prelim}
	\subsection{The EHR-based two-phase sampling framework\label{sec:framework}}
\begin{figure}[h]
    \centering
    \includegraphics[width=\linewidth]{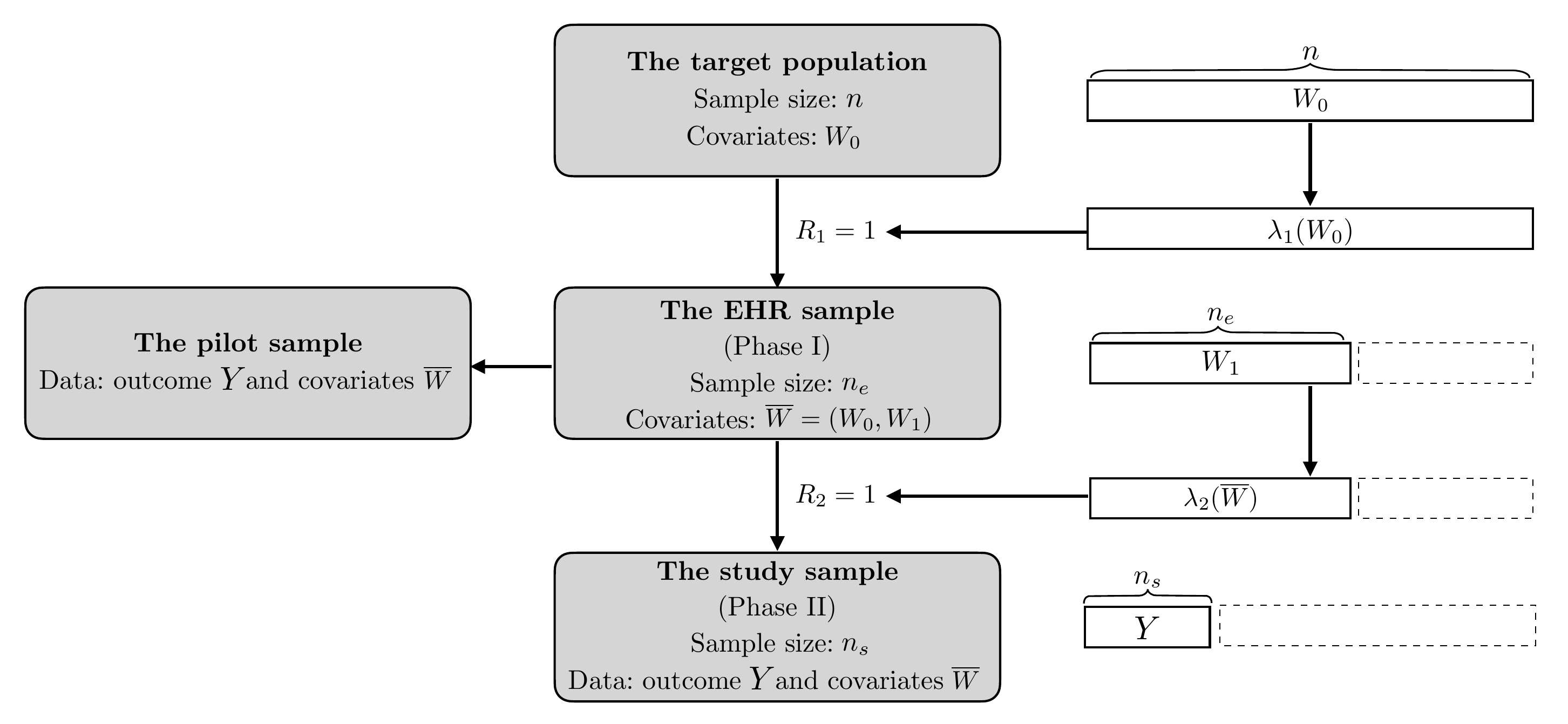}
    \caption{Flowchart of the EHR-based two-phase sampling framework. 
    We aim to recruit a study sample based on a designed and optimized sampling probability $\lambda_{2}(\overline{W})$ to minimize the variance of the estimated parameter. 
	The dashed boxes indicate missing data, and the solid boxes indicate observed data at each phase.  
 }
    \label{fig:flowchart}
\end{figure}

	Let $Y$ denote the outcome of interest that is expensive or time-consuming to measure. Suppose one is interested in estimating the mean outcome, $\beta=E[Y]$, in the target population. This can be generalized to other parameters of interest, such as the ATE and regression coefficients, which we detail in Section~\ref{sec:ext}. Our goal is to develop an optimal sampling mechanism to improve efficiency in \textcolor{black}{the} estimation of $\beta$ by leveraging the auxiliary information that is inexpensive and readily available in the EHR sample. 
	Below we summarize the samples at each phase, the relationship between the samples, and the data missingness pattern, which are visualized in Figure~\ref{fig:flowchart}. 

\paragraph*{The target population}
	Consider a target population of $n$ individuals. Let $W_0$ denote patient characteristics predictive of both the outcome and whether a patient seeks healthcare and thus show\textcolor{black}{s} up in the EHR system. 
	We assume that information about $W_0$ is available in either of the following 
	two scenarios: 
	(1) $W_0$ is available in an external probability sample, such as a survey study, in which individuals are sampled from the target population with known sampling mechanism; (2) the distribution (or at least summary statistics) of $W_0$ in the target population is available from a public data source, such as the Census.

	\paragraph*{The {\phaseone} sample: the EHR sample} 
	Suppose one has access to EHR data on 
	a subset of the target population with sample size $n_e\leq n$, referred to as the EHR sample or the {\phaseone} sample.
	Let $R_{1}\in\{0,1\}$ be the indicator of whether an individual in the target population is selected into the {\phaseone} sample, then $n_e=\sum_{i=1}^n R_{1i}$. 
	The mechanism of being selected into the {\phaseone} sample, i.e., the probability of having EHR available for an individual in the target population, depends on  patient characteristics $W_0$ via an unknown selection mechanism 
	\begin{equation*}
		\lambda_{1}(W_0) = P(R_{1}=1\mid W_0).
	\end{equation*}
	Let $W_1$ denote auxiliary covariates that are available in {\phaseone} in addition to $W_0$ and \textcolor{black}{are} predictive of the outcome. 
	To ease exposition, let $\overline{W}=(W_0,W_1)$ denote all covariates available in the {\phaseone} sample, that is, $\overline{W}_{i}$ is observed if $R_{1i}=1,i=1,\dots,n$. 
	
	\paragraph*{The {\phasetwo} sample: the study sample} The study sample of size $n_s$ $(1\leq n_s\leq n_e)$, also referred to as the {\phasetwo} sample, will be recruited from the {\phaseone} EHR sample based on patient characteristics $\overline{W}$. Then a clinical outcome of interest, $Y$, will be measured, which often involves labor-intensive manual chart review in EHR-based research. 
	Let $R_{2}\in\{0,1\}$ be the indicator of whether an individual is recruited into the {\phasetwo} sample, then $n_s=\sum_{i=1}^n R_{2i}$. 
	The probability of being selected into the {\phasetwo} sample from the {\phaseone} sample depends on patient characteristics $\overline{W}$ via a designed probability
	\begin{equation*}
		\lambda_{2}(\overline{W}) = P(R_{2}=1\mid \overline{W},R_{1}=1).
	\end{equation*}
 
{\color{black}To summarize, ideally we wish to have i.i.d. sample of the complete data $(Y, W_1,W_0)$, which arises from some joint distribution $P\in\mathcal{M}$ in the target population. Sampling into EHR generates incomplete data $(W_1R_1,W_0)\sim P_{\scriptscriptstyle\text{I}}\in \mathcal{M}_{\scriptscriptstyle\text{I}} = \{P_{\scriptscriptstyle{P,\lambda_1(W_0)}}: P \in \mathcal{M}, \lambda_1(W_0)\}$, where model $\mathcal{M}_{\scriptscriptstyle\text{I}}$ is implied by biased sampling through $\lambda_1(W_0)$ (i.e., phase-I sampling) from $\mathcal{M}$. To study $\beta=E[Y]$, one then draws a phase-II sample to measure the outcome  through $\lambda_2(\overline{W})$, which generates the final observed data 
 $O=(YR_2,W_1R_1,W_0)\sim P_{\scriptscriptstyle\text{II}}\in \mathcal{M}_{\scriptscriptstyle\text{II}} = \{P_{\scriptscriptstyle{P_{\scriptscriptstyle\text{I}},\lambda_2(\overline{W})}}: P_{\scriptscriptstyle\text{I}}\in \mathcal{M}_{\scriptscriptstyle\text{I}}, \lambda_2(\overline{W})\}$.} We aim to find the optimal sampling design, denoted as $\lambda_{2}^*(\overline{W})$, 
	that is most efficient in the sense that the sampling design minimizes the asymptotic variance for a given estimator of $\beta$.

	\paragraph*{The pilot sample} Design of a study typically relies on either certain prior knowledge or preliminary data obtained from a pilot study. 
	We assume that a small pilot sample of size $n_p$ is available with both $Y$ and $\overline{W}$ measured. The pilot data allow us to model the relationship between the outcome and the auxiliary covariates to inform {\phasetwo} sampling design.
	Ideally, the pilot sample should be a random sample of the {\phaseone} EHR sample.  When such a pilot sample is unavailable, knowledge about the conditional variance of the outcome, $Var(Y\mid \overline{W},R_1=1)$,  is required.

	In addition to the multi-phase samples, we introduce some notation for budget consideration.
	Let $B$ denote the total budget, $C_0$ denote the initial study cost, and $C_1$ denote the per-individual cost in {\phaseone} that scales with the EHR sample size
	and does not depend on patient characteristics. Let $C_2(\overline{W})$ be the per-individual cost in {\phasetwo} that may depend on auxiliary covariates for patient characteristics. 
	\textcolor{black}{For example, in HIV vaccine trials, the cost of measuring immune response is associated with the number of reactive HIV epitopes \citep{zolla2004identifying,sahay2017conserved}.} The total expected cost, which is constrained to not exceed the total budget $B$, is given by 
	$C_0+n_eC_1+nE\{\lambda_{1}(W_0)\lambda_{2}(\overline{W})C_2(\overline{W})\}$, which is derived in Section 2.1 of the Supplementary Material \citep{zhang2022supp}.

\subsection{Estimation under two-phase sampling allowing for a biased {\phaseone} sample\label{subsec:biased}}
	\citet{gilbert2014optimal} proposed an optimal {\phasetwo} sampling design that minimizes the  variance of an estimator of $\beta$ with a budget constraint, under the assumption that the {\phaseone} sample is a simple random sample of the target population, that is, $W_0=\emptyset$ and $\lambda_{1}(W_0)$ is a constant. In our setting, the {\phaseone}  sample is a cohort of patients whose EHR data are available to the investigator. The {\phaseone} EHR sample is not necessarily a random sample representative of the target population, but rather is potentially subject to  selection bias, which needs to be accounted for.
	In this section, we extend the optimal sampling design proposed by \citet{gilbert2014optimal} to account for  selection bias in the {\phaseone} EHR sample with an unknown  mechanism $\lambda_1(W_0)$. 
	
	We impose the following standard assumptions in the missing data literature: 
	\begin{itemize}
	    \item[] 1. Missing at random: $R_2\ind Y \mid R_1,\overline{W}$ and $R_1\ind (Y, W_1) \mid W_0$;
	    \item[] 2. {\color{black}Positivity: $\epsilon < \lambda_1(W_0), \lambda_2(\overline{W}) \leq 1$ for some $\epsilon > 0$;}
	    \item[] 3. One of the following two sets of models is correctly specified:
	    \begin{enumerate}
	        \item[] (i) Selection probabilities: $\lambda_{1}(W_0)$ and $\lambda_{2}(\overline{W})$;  
	        \item[] (ii) Outcome regression model: $E(Y\mid \overline{W},R_2=1)$ and $f(W_1\mid W_0,R_1=1)$ (thus $E(Y\mid W_0,R_1=1)$ is also correctly specified).
	    \end{enumerate}
	\end{itemize}


	Assumption 1 implies that in each phase, the sampling mechanism only depends on some auxiliary covariates of the previous phase and does not depend on the outcome and auxiliary covariates of the next phase. Specifically, in {\phaseone}, we assume that the sampling probability is an unknown function of $W_0$, 
	which is a more flexible setting than the traditional two-phase sampling framework proposed by \citet{gilbert2014optimal} where $\lambda_{1}(W_0)$ is assumed to be a constant. 
	In {\phasetwo}, Assumption 1 is guaranteed to hold because we will sample patients based on the derived optimal sampling probability $\lambda^*_2(\overline{W})$, which does not depend on $Y$. 
	In the motivating example of patient recruitment through MGI, auxiliary covariates predictive of the outcome and available in MGI data can serve as $\overline{W}$ in practice. Among $\overline{W}$, covariates with a differential distribution between the MGI sample and the target population should be taken as $W_0$ to standardize results to the target population.

	Assumption 2 requires that individuals with any given value of $W_0$ (or $\overline{W}$) have nonzero probabilities of being selected into the EHR sample (or study sample). Positivity of $\lambda_{2}(\overline{W})$ holds by design. 
{\color{black}The positivity assumption of $\lambda_{1}(W_0)$ is more stringent. For example, suppose health insurance coverage impacts whether a patient seeks care at the University of Michigan Health System, such that there is zero probability for patients without insurance to show up in MGI data. Then the positivity assumption requires that health insurance coverage as a covariate is not needed for Assumption 1 to hold, that is, access to insurance does not impact the disease outcome. Under a weaker version of Assumption 1 referred to as the mean exchangeability \citep{shi2023data}, i.e., $E[Y\mid W_0, R_1=1]=E[Y\mid W_0]$, the positivity assumption requires that health insurance coverage does not modify the outcome model in the MGI sample compared to the target population.
In practice, one should carefully evaluate factors that might impact both the outcome of interest and the selection into the EHR sample before applying our proposed method.} 

	Assumption 3 is imposed to ensure consistent estimation of $\beta$, as detailed in Section 2.2 of the Supplementary Material \citep{zhang2022supp}.
	
	As illustrated in Figure~\ref{fig:flowchart}, the EHR-based two-phase sampling framework constitutes a monotone missing data pattern, with independent and identically distributed observations $O_i=(Y_iR_{2i},W_{1i}R_{1i}, W_{0i})$, $i=1,\dots,n$. Under Assumptions 1-3,  \citet{rotnitzky1995semiparametric} proposed to estimate $\beta$ by solving $n^{-1}\sum_{i=1}^n U(O_i;\beta)=0$, where 
	\begin{equation} \label{eq:influence}
	\begin{split}
	    U(O;\beta)&=
			\frac{R_1R_{2}}{\lambda_{1}(W_0)\lambda_{2}(\overline{W})} Y
			-\frac{R_1R_{2}-R_1\lambda_{2}(\overline{W})}{\lambda_{1}(W_0)\lambda_{2}(\overline{W})}E(Y\mid \overline{W},R_{1}=1)\\
			&-\frac{R_{1}-\lambda_{1}(W_0)}{\lambda_{1}(W_0)}E(Y\mid W_0)-\beta.
	\end{split}
	\end{equation}
 We refer to the resulting estimator as the RR estimator hereafter \citep{rotnitzky1995semiparametric}.
	The estimating function $U(O;\beta)$ is originally derived for mean outcome estimation in the presence of nonresponse in longitudinal studies. The estimation strategy directly applies to our setting where the sampling mechanisms in multiple phases (see Figure~\ref{fig:flowchart}) correspond to the nonresponse mechanisms over time. 
	Intuitively, one could estimate $\beta$ while accounting for biased sampling by inverse probability weighting using the measured outcome from the {\phasetwo} sample, which correspond\textcolor{black}{s} to the first term, ${R_1R_{2}Y}/\{\lambda_{1}(W_0)\lambda_{2}(\overline{W})\}$.
	Efficiency can be improved by further incorporating information from the auxiliary covariates through two augmentation terms that essentially imputes  missing outcomes using $E(Y\mid W_0)$ and $E(Y\mid \overline{W},R_{1}=1)$ when the outcome models are correctly specified. 

 {\color{black}The estimator is doubly robust in the sense that $E(U(O;\beta))=0$ when selection probabilities (i.e., $\lambda_{1}(W_0)$ and $\lambda_{2}(\overline{W})$) or outcome models (i.e., $E(Y\mid \overline{W},R_1=1)$ and $E(Y\mid W_0)$) are correctly specified, which is proved in Section 2.2 of the Supplementary Material. Based on Assumption (1), we have 
		\begin{align*}
			&E(Y\mid \overline{W},R_1=1)=E(Y\mid \overline{W},R_2=1)\text{ and}\\
			&E(Y\mid W_0)=E(Y\mid W_0,R_1=1)=E\{E(Y\mid \overline{W},R_2=1)\mid W_0,R_1=1\}.
		\end{align*}
Therefore,  correct specification of $E(Y\mid \overline{W}, R_1=1)$ is equivalent to  correct specification of $E(Y\mid \overline{W}, R_2=1)$. Similarly,  correct specification of $E(Y\mid W_0)$ is equivalent to  correct specification of $E(Y\mid \overline{W}, R_2=1)$ and $f(W_1\mid W_0, R_1=1)$.
Thus consistent estimation is  ensured when one of the following two sets of models is correctly specified:
	    \begin{enumerate}
	        \item[] (i) Selection probabilities: $\lambda_{1}(W_0)$ and $\lambda_{2}(\overline{W})$;  
	        \item[] (ii) Outcome regression model: $E(Y\mid \overline{W},R_2=1)$ and $f(W_1\mid W_0,R_1=1)$.
	    \end{enumerate}}

	The asymptotic variance of the RR estimator is given by $V(\lambda_2)/n$, where
	\begin{equation*}\label{eq:variance}
\begin{split}
    V(\lambda_2)&=Var(Y) + E\left[\left\{\frac{1}{\lambda_1(W_0)}-1\right\}Var(Y\mid W_0)\right]\\
    &+ E\left[\left\{\frac{1}{\lambda_2(\overline{W})}-1\right\}\frac{1}{\lambda_1(W_0)}Var(Y\mid \overline{W},R_1=1)\right].
\end{split}
	\end{equation*}

\section{Methods\label{sec:method}}
	\subsection{Optimal two-phase sampling allowing for a biased {\phaseone} sample\label{subsec:opt}}
 {\color{black}We aim to find the optimal sampling probability $\lambda^*_2(\overline{W})$ that minimizes the above asymptotic variance of the RR estimator under the constraint that the phase-II sampling probability is non-negative and that the  budget covers the total expected cost. This can be formulated as the following optimization problem:
 \begin{equation*}
     \begin{split}
         \arg\min_{\lambda_2(\overline{w})}\;\;  &\frac{V(\lambda_2)}{n}  \\
\text{s.t.}\;\;  &0<\lambda_2(\overline{W})\leq1 \\
  &C_0+n_eC_1+nE\{\lambda_{1}(W_0)\lambda_{2}(\overline{W})C_2(\overline{W})-B =0,
     \end{split}
 \end{equation*}
	 which is a convex optimization problem that can be solved with Karush–Kuhn–Tucker (KKT) conditions.}
	The rationale behind using the KKT conditions is the following. 
	To find the extremum (maximum or minimum value) of a function in an unconstrained optimization problem, one usually searches for an optimal point where the slope or gradient is zero. When the optimization problem is subject to an equality constraint, the use of the Lagrange multiplier helps convert an optimization problem into a system of equations. The solution to the system of equations is the optimal point. The KKT method generalizes the method of Lagrange multipliers to inequality constraints. 
     It has been shown that for a convex optimization problem, the point that satisfies the KKT conditions is the sufficient and necessary solution for optimality \citep{boyd2004convex}.
     
	Given a fixed budget $B$, target population sample size $n$, and EHR sample size $n_e$, we now present the optimal sampling probability $\lambda^*_2(\overline{W})$ that satisfies the KKT conditions and hence achieves the minimal variance, which is proved in  Section 2.3 of the Supplementary Material \citep{zhang2022supp}.
	\begin{theorem}\label{thm:fixed_n}
		For fixed $B$ and $n_e$ that satisfy $n_e<(B-C_0)/C_1$,
		the minimal variance among all possible designs that do not exceed the budget $B$ is achieved at
		\begin{equation}\label{eq:lambda_thm1}
				\lambda^*_2(\overline{W}=\overline{w}) =  \min \left\{1, \frac{B-C_0-n_eC_1}{n\lambda_1(w_0)}\frac{\sqrt{{Var(Y\mid \overline{W},R_1=1)}/{C_2(\overline{w})}}}{E\left\{\sqrt{C_2(\overline{W})Var(Y\mid \overline{W},R_1=1)}\right\}}\right\}.
		\end{equation}
	\end{theorem}
	The upper bound for $n_e$ is to make sure that we have enough budget to cover the per-individual cost. We can see that $\lambda^*_2(\overline{W})$ increases with $Var(Y\mid \overline{W},R_1=1)/C_2(\overline{W})$, which is the cost-standardized conditional variance of $Y$.
	Intuitively, individuals with noninformative auxiliary covariates (i.e., $Var(Y\mid \overline{W},R_1=1)$ is large) and relatively affordable outcome measurement costs (i.e., $C_2(\overline{W})$ is low) will be oversampled by the proposed design, because for these individuals, it might be more efficient to measure their outcome $Y$ directly. Conversely, individuals with informative auxiliary covariates and expensive outcome measurement costs will be undersampled, because for these individuals, instead of directly measuring the outcome, it is more efficient to impute the unobserved outcome leveraging the highly predictive auxiliary covariates.  
	
	It is important to note  that $Var(Y\mid \overline{W},R_1=1)$ and $\lambda_1(W_0)$  in Eq. (\ref{eq:lambda_thm1}) are generally unknown and need to be estimated to inform {\phasetwo} sampling. We propose to estimate $Var(Y\mid \overline{W},R_1=1)$ using data from the pilot sample, which is detailed in  Section 2.4  of the Supplementary Material \citep{zhang2022supp}. When individual-level data on $W_0$ are available in the target population, it is straightforward to estimate $\lambda_1(W_0)$ by running a regression model for $P(R_1=1\mid W_0)$. However, in practice, we generally do not have individual-level data from the target population. We address this issue in the next section.
	
	\subsection{Addressing selection bias in the {\phaseone} EHR sample\label{subsec:bias}}
	In this section, we present two methods to account for potential selection bias $\lambda_1(W_0)$ in the EHR sample: direct estimation of $\lambda_1(W_0)$ and indirect bias reduction via subsampling. 
	
	\subsubsection{Method 1: direct estimation of selection mechanism\label{sec:lambda1_method1}}
	\citet{Beesley2019.12.26.19015859} proposed novel strategies to model $\lambda_1(W_0)$ by leveraging an external probability sample of the target population with both sampling probability and the auxiliary covariates available, assuming there is no overlap between the external probability sample and the EHR sample. 
	Such external data are often available in survey studies. For example, one can use the publicly available National Health and Nutrition Examination Survey (NHANES) data as the external probability sample when the target population is defined as the US adult population. The rationale is to approximate the {\phaseone} sampling probability by calibrating the sampling probability of the external probability sample (such as NHANES) \citep{elliot2013combining}.
	
	We employ the 
	method proposed by \cite{Beesley2019.12.26.19015859} to estimate the probability of being selected into the {\phaseone} EHR sample. 
	Specifically,  we have that
	\begin{equation}\label{eq:lambda1}
		\lambda_1(W_0)=P(R_1=1\mid W_0)\approx P(R_{\scriptscriptstyle \text{prob}}=1\mid W_0)\frac{P(R_1=1\mid W_0,R_{\scriptscriptstyle \text{comb}}=1)}{1-P(R_1=1\mid W_0,R_{\scriptscriptstyle \text{comb}}=1)},
	\end{equation}
	where $R_{\scriptscriptstyle \text{prob}}$ is the indicator of being included in the external probability sample from the target population, $R_{\scriptscriptstyle \text{comb}}$ is the indicator of being included in the sample combining both the EHR sample and the external probability sample, that is, $R_{\scriptscriptstyle \text{comb}}=\mathbb{I}(R_{\scriptscriptstyle \text{prob}}=1\text{ or }R_1=1)$.
 
	 \textcolor{black}{To estimate $P(R_{\scriptscriptstyle \text{prob}}=1\mid W_0)$, we take the sampling probability in the external probability sample as a continuous outcome in $[0, 1]$ and regress it on the auxiliary covariates $W_0$ by fitting a regression model, the most common choices are beta regression and simplex regression \citep{kieschnick2003regression}. Beta regression is a natural choice when one believes that the outcome follows a beta distribution, while simplex regression is used if the empirical distribution of the sampling probability resembles a bimodal pattern \citep{zhang2014regression,espinheira2018nonlinear}.}
	We estimate $P(R_1=1\mid W_0,R_{\scriptscriptstyle \text{comb}}=1)$ using the combined data consisting of both the EHR sample and the \textcolor{black}{external probability sample} by fitting a generalized linear model. 
	Finally, we estimate $\lambda_1(W_0)$, the probability of being selected into the {\phaseone} EHR sample, via Eq. (\ref{eq:lambda1}).

	
	\subsubsection{Method 2: indirectly addressing selection bias by subsampling strategy\label{sec:lambda1_method2}} 
 Instead of directly estimating $\lambda_1(W_0)$, an alternative approach to address selection bias is to draw an unbiased subsample from the biased {\phaseone} EHR sample, such that the joint distribution (or at least summary statistics) of $W_0$ in this subsample is the same as that of the target population.
	We refer to this subsample as {the new {\phaseone} sample}, which can be treated as a random sample of the target population. 
	A practical approach to obtain such an unbiased subsample from the biased EHR sample is through matching. {\color{black}Specifically, we first simulate a sample of $W_0$'s according to the distribution  of $W_0$ in the target population, then we match the biased EHR sample to the unbiased $W_0$ data using the nearest neighbor matching method without replacement  \citep{ho2007matching,stuart2010matching}. 
This matching method coincides with the idea of ``template matching'' proposed by \cite{bennett2020building}.

The  new {\phaseone} sample selected by matching is a random sample of the target population. 
 Therefore, the methods proposed by  \cite{gilbert2014optimal} immediately apply. Following our notation and study settings, below we provide the optimal {\phasetwo} sampling design under subsampling, which is {analogous} to Result 3 in \cite{gilbert2014optimal}.}
	
	Let $R_{1}'\in\{0,1\}$ be the indicator of whether an individual in the target population is selected into the new {\phaseone} sample. Let $n_e' ~(n_e'\leq n_e)$ denote the sample size of the new {\phaseone} sample, then the probability of being selected into the new {\phaseone} sample from the target population is a constant given by ${\lambda^{\scriptscriptstyle \text{alt}}_1}=n_e'/n$.
	Similar to Theorem \ref{thm:fixed_n}, the optimal   {\phasetwo} sampling probability under the new {\phaseone} sample is
	\begin{equation}\label{eq:lambda_col}
		\lambda^{{\text{alt}}}_2(\overline{W}=\overline{w}) = \min\left\{1,\frac{B-C_0-n_e'C_1}{n{\lambda^{\scriptscriptstyle \text{alt}}_1}}\frac{\sqrt{{Var(Y\mid \overline{W},R_1'=1)}/{C_2(\overline{w})}}}{E\left\{\sqrt{C_2(\overline{W})Var(Y\mid \overline{W},R_1'=1)}\right\}}\right\},
	\end{equation}
where $Var(Y\mid \overline{W},R_1'=1)=Var(Y\mid \overline{W})$. We refer to the optimal {\phasetwo} sampling probability under a new {\phaseone} sample as the {alternative two-phase sampling design}. 
	
	\subsection{Mean outcome estimation\label{subsec:estbeta}}
	After the {\phasetwo} study sample is recruited and the outcome is measured, we  estimate $\beta$ via the RR estimator \citep{rotnitzky1995semiparametric}
		\begin{equation*} \label{eq:beta}
		\begin{split}
			{\color{black}\widehat{\beta}_{RR}}=\frac{1}{n}\sum_{i=1}^{n}\Bigg\{\frac{R_{1i}R_{2i}}{\widehat{\lambda}_{1i}\lambda_{2i}}Y_i - \frac{R_{1i}R_{2i}-R_{1i}\lambda_{2i}}{\widehat{\lambda}_{1i}\lambda_{2i}}\widehat{E}(Y\mid \overline{W}_{i},R_{1i}=1) - \frac{R_{1i}-\widehat{\lambda}_{1i}}{\widehat{\lambda}_{1i}}\widehat{E}(Y\mid W_{0i})\Bigg\}.
		\end{split}
	\end{equation*}
 Note that $R_1$ should be replaced with $R_1'$ if method 2 was used.

 A key step is to estimate the outcome models $E(Y\mid W_0)$ and $E(Y\mid \overline{W},R_1=1)$.  Recall that by Assumption (1), we have $E(Y\mid W_0)=E\{E(Y\mid \overline{W},R_2=1)\mid W_0,R_1=1\}$ and $E(Y\mid \overline{W},R_1=1)=E(Y\mid \overline{W},R_2=1)$. Thus these models can be estimated using outcomes measured in  the {\phasetwo}  sample or from some existing pilot data. 
		We detail the estimation strategies in  Section 2.4  of the Supplementary Material \citep{zhang2022supp}.
	 \textcolor{black}{A summary of the overall procedure for sampling design and estimation is presented in Figure~\ref{fig:algorithm}. We first estimate $Var(Y\mid \overline{W},R_1=1)$ and $\lambda_1(W_0)$ to obtain the optimal phase-II sampling design, $\lambda_2^*(\overline{W})$,  based on Eq. (\ref{eq:lambda_thm1}). Then we draw the phase-II sample based on $\lambda_2^*(\overline{W})$ and measure the expensive outcome for each individual in the phase-II sample. Finally, we estimate the parameter based on the proposed RR estimator.}
	
\begin{figure}
		\centering
		\tikzstyle{block} = [rectangle, draw, fill=gray!20, 
		text width=6.5 cm, text centered, 
		rounded corners, minimum height=0.75cm]
		\tikzstyle{wideblock} = [rectangle, draw, fill=gray!20, text centered, 
		text width=15 cm, rounded corners, minimum height=0.75cm]
		\tikzstyle{midblock} = [rectangle, draw, fill=gray!20, text centered, 
		text width=8 cm, rounded corners, minimum height=0.75]
		\tikzstyle{line} = [draw, -latex']
		\tikzstyle{vecArrow} = [thick, decoration={markings,mark=at position
			1 with {\arrow[semithick]{triangle 60}}},
		double distance=1.4pt, shorten >= 5.5pt,
		preaction = {decorate},
		postaction = {draw,line width=1.4pt, white,shorten >= 4.5pt}]
		\tikzstyle{innerWhite} = [semithick, black,line width=1.4pt, shorten >= 4.5pt]
		\scalebox{0.8}{\begin{tikzpicture}[node distance = 5cm, auto]
			Place nodes
			\node[wideblock](inp){\textbf{Input}\vspace{-0.4cm} \begin{flushleft}\begin{multicols}{3}$Y$: outcome variable \\ $\overline{W}$: auxiliary covariates \\ $n$: the size of the target population \\
						$n_e$: the size of the EHR sample\\ $n_p$: the size of the pilot sample \\ $B$: total budget \\ $C_0$: the initial study cost \\$C_1$: the per-individual cost \\ $C_2(\overline{W})$: the auxiliary-specific cost  \end{multicols}\end{flushleft}};
			\node[wideblock, below=1cm of inp, yshift=-1cm](l2){\textbf{Calculate $\boldsymbol{\lambda^*_2(\overline{W})}$} with ${Var(Y\mid \overline{W}, R_1=1)}$ and $\lambda_1(W_0)$};
			\node[block, left=-6.8cm of l2, xshift=-0.1cm, yshift=1.5cm](var){\textbf{Estimate $\boldsymbol{Var(Y\!\mid\! \overline{W}, \!R_1\!=\!1)}$} using the pilot sample with $Y$ available};
			\node[block, right=-6.8cm of l2, xshift=0.1cm, yshift=1.5cm](l1){\textbf{Estimate $\boldsymbol{\lambda_1(W_0)}$} directly or  address selection bias in the EHR sample indirectly};
			\node[wideblock, below=0.5cm of l2, yshift=0cm](p2){\textbf{Draw the {\phasetwo} sample} according to $\lambda^*_2(\overline{W})$, measure $Y$};
			\node[wideblock, below=0.5cm of p2, yshift=0cm](e){\textbf{Estimate $\boldsymbol{E(Y\mid W_0)}$ and $\boldsymbol{E(Y\mid \overline{W},R_1=1)}$} using either the {\phasetwo} sample or the pilot sample};
			\node[wideblock, below=0.5cm of e, yshift=0cm](beta){\textbf{Estimate $\boldsymbol{\beta}$} using $Var(Y\!\mid\! \overline{W}, \!R_1\!=\!1)$, $\lambda_1(W_0)$, $\lambda^*_2(\overline{W})$, ${E(Y\mid W_0)}$, and ${E(Y\mid \overline{W},R_1=1)}$};
			\node[midblock, below=0.5cm of beta, yshift=0cm](out){\textbf{Output}\\ $\lambda^*_2(\overline{W})$: the  {\phasetwo} sampling probability\\  {\color{black}$\widehat{\beta}_{RR}$}: the estimated mean outcome\\
            {\color{black}$\widehat{V}$: the estimated variance of the estimated mean outcome}};
			\path [line] (var) -- (l2);
			\path [line] (l1) -- (l2);
			\path [line] (l2) -- (p2);
			\path [line] (p2) -- (e);
			\path [line] (e) -- (beta);
			
			\draw[vecArrow] (var) -- (l2);
			\draw[vecArrow] (l1) -- (l2);
			\draw[vecArrow] (l2) -- (p2);
			\draw[vecArrow] (p2) -- (e);
			\draw[vecArrow] (e) -- (beta);
			
			\draw[innerWhite] (var) -- (l2);
			\draw[innerWhite] (l1) -- (l2);
			\draw[innerWhite] (l2) -- (p2);
			\draw[innerWhite] (p2) -- (e);
			\draw[innerWhite] (e) -- (beta);
		\end{tikzpicture}}
		
		\caption{General Procedure for Efficient Two-Phase Sampling and Mean Estimation.\label{fig:algorithm}}
	\end{figure}

\section{Relative efficiency comparing optimal two-phase sampling and random sampling\label{sec:RE}}
	In this section, we show that our proposed optimal two-phase sampling design is more efficient than random sampling. 
	We define the relative efficiency (RE) comparing the asymptotic variance of the RR estimator obtained using data from optimal {\phasetwo} sampling to that of the RR estimator but using data from random {\phasetwo} sampling as 
	$$\text{RE}=\frac{V\{ \lambda^*_2(\overline{W}) \}}{V(\overline{\lambda}_2)},$$ where $\overline{\lambda}_2 = (B-C_0-n_eC_1)/[n\lambda_1(W_0)E\{C_2(\overline{W})\}]$ is the random sampling probability derived under the budget constraint. 
	The RE measures the efficiency gain purely due to the proposed optimal auxiliary-dependent sampling design. 
	When $\text{RE} \leq1$, $V\{ \lambda^*_2(\overline{W})\}\leq V(\overline{\lambda}_2)$.  
	The smaller $\text{RE}$ is, the more efficiency gain we obtain from the optimal sampling.  
	Further define the \underline{p}roportion of the \underline{v}ariation in the outcome \underline{e}xplained (PVE) by covariates $\overline{W}$ in the {\phaseone} EHR sample as $\text{PVE}=Var\{E(Y\mid \overline{W},R_1=1)\}/Var(Y)$. The PVE  measures how  \textcolor{black}{ well $\overline{W}$ predicts the outcome $Y$}.
	Below we present  RE under the two  methods proposed in Section~\ref{subsec:bias} for addressing selection bias in the {\phaseone}  sample.
 
	\begin{corollary} \label{cor:total}~\\\vspace{-\baselineskip}
 
		(i) Suppose $B$ and $n_e$ satisfy $n_e<(B-C_0)/C_1$.
		If selection bias is addressed via direct estimation of selection mechanism (Section~\ref{sec:lambda1_method1}), the relative efficiency comparing the proposed optimal sampling to random {\phasetwo} sampling is
		\begin{equation*}
			\text{RE}=\frac{\text{PVE}*Var(Y) + E'\{\lambda_1(W_0)\}+  E\left\{\frac{1}{\lambda^*_2(\overline{W})\lambda_1(W_0)}Var(Y\mid \overline{W},R_1=1)\right\}}{\text{PVE}*Var(Y) + E'\{\lambda_1(W_0)\}+  E\left\{\frac{1}{\overline{\lambda}_2\lambda_1(W_0)}Var(Y\mid \overline{W},R_1=1)\right\}}
			\leq 1,
		\end{equation*}
		where 
		$E'\{\lambda_1(W_0)\}=E\left[(1-\lambda_1(W_0))\Big\{Var(Y\mid W_0)-Var(Y\mid \overline{W},R_1=1)\Big\}/\lambda_1(W_0)\right]$. 
  
  {\color{black}(ii) If selection {bias} is addressed indirectly via subsampling (Section \ref{sec:lambda1_method2}), there is no clear pattern of $\text{RE}={V\{ \lambda^{{\text{alt}}}_2(\overline{W})\}}/{V(\overline{\lambda}_2)}$ where $\overline{\lambda}_2$ is random phase-II sampling from the original phase-I EHR sample. Nevertheless, the relative efficiency comparing the optimal phase-II sampling to random phase-II sampling (both sampling from the new phase-I sample)
  \[\text{RE}^{{\text{alt}}}=\frac{V\{\lambda^{{\text{alt}}}_2(\overline{W})\}}{V(\overline{\lambda})}\leq 1\]
  when $n_e'<(B-C_0)/C_1$, where $\overline{\lambda} = (B-C_0-n_e'C_1)/[n_e'E\{C_2(\overline{W})\}]$ is the random phase-II sampling from the new phase-I sample under the budget constraint.  
}
  \end{corollary}

  	Corollary~\ref{cor:total}  is proved in  Section 2.5  of the Supplementary Material \citep{zhang2022supp}. Corollary~\ref{cor:total} (i) states that our proposed optimal sampling design improves efficiency compared to  random sampling design. {\color{black} Corollary~\ref{cor:total} (ii) states that if one starts from the new phase-I sample of size $n_e'$ (ignoring the additional $n_e-n_e'$ data points in the original phase-I EHR sample), which is a random sample from the target population, then optimal sampling improves efficiency compared to random sampling. This conclusion about $\text{RE}^{{\text{alt}}}$ is {analogous}  to Eq. (11) of \cite{gilbert2014optimal}.} We also found that RE and RE$^\text{alt}$ increase with PVE, which indicates that we obtain less efficiency gain from the optimal sampling design when PVE is large, i.e., the auxiliary covariates $\overline{W}$ explains a major portion of the variation in the outcome in the {\phaseone} EHR sample. Intuitively, if $\overline{W}$ is sufficiently informative in predicting the outcome, then efficiency can be largely achieved by estimation with outcome imputation alone. 
	As such, efficiency gain due to sampling is less obvious when the  predictive power of $\overline{W}$  is higher.

{\color{black} Although there is no clear pattern of the relative efficiency comparing ${V\{ \lambda^{{\text{alt}}}_2(\overline{W})\}}$ and ${V(\overline{\lambda}_2)}$, we study how it increases or decreases with $n_e'$  in Section 2.5 of the Supplementary Material.
We have also provided Table C1 in Section 2.9 of the Supplementary Material to summarize and clarify the difference between sampling methods mentioned in this paper.
}	
	
	\section{Extension to a general design framework\label{sec:ext}}
		Our proposed two-phase sampling design suggests a general auxiliary-covariate-based sampling framework to improve efficiency: for a given estimator which is the solution to an
		estimating equation that incorporates the selection mechanism in each phase, one derives the optimal {\phasetwo} sampling probability by minimizing the asymptotic variance of the estimator under a budget constraint. 
		Based on this idea, in this section, we extend our methods to the estimation of 
		a general parameter of interest, $\beta$, defined as the unique solution of an estimating equation $E\{m(Y,\overline{W};\beta)\}=0$. 
		
		When $(Y,\overline{W})$ is observed on a random sample of the target population, it is straightforward to estimate $\beta$ by solving $\sum_{i} m(Y_i,\overline{W}_i;\beta)=0$. 
		In contrast, the EHR-based two-phase sampling framework constitutes a monotone missing data pattern with observations $O_i=(Y_iR_{2i},W_{1i}R_{1i}, W_{0i})$, $i=1,\dots,n$, as discussed in Section~\ref{subsec:biased}. In this setting, 
		 \citet{tsiatis2006semiparametric} presented an 
		 augmented inverse probability weighted complete-case estimator for $\beta$ that solves $E[U(O;\beta)]=0$, where
		\begin{equation}\label{eq:generalif} 
	\begin{split}
	    U(O;\beta)&=
			\frac{R_1R_{2}}{\lambda_{1}(W_0)\lambda_{2}(\overline{W})} m(Y,\overline{W};\beta)
			-\frac{R_1R_{2}-\lambda_{2}(\overline{W})R_{1}}{\lambda_{1}(W_0)\lambda_{2}(\overline{W})}E\{m(Y,\overline{W};\beta))\mid \overline{W},R_{1}=1\}\\
			&-\frac{R_{1}-\lambda_{1}(W_0)}{\lambda_{1}(W_0)}E\{m(Y,\overline{W};\beta)\mid W_0\}.
	\end{split}
	\end{equation}
	One can then derive the optimal sampling probability $\lambda_{2}(\overline{W})$ based on our proposed framework. We now illustrate this framework with two examples: (1) estimation of causal effects in observational studies or RCTs, and (2) estimation of regression coefficients.
		
	\subsection{Design  and estimation for causal inference\label{subsec:ate}}
	Following the potential outcome framework \citep{neyman1923application,rubin1974estimating,rubin2005}, we define a pair of potential outcomes $(Y(1),Y(0))$, representing the outcomes had an individual received treatment or control.
	Consider a binary treatment $A$, with $A=1$ if an individual received treatment and $0$ otherwise.
	Under the standard assumption of consistency, we observe outcome $Y=Y(a)$ if $A=a$, for $a=0,1$. 
	Let $\beta_a=E\{Y(a)\}$, $a=0,1$, denote the mean potential outcome.  The ATE is a contrast between $\beta_1$ and $\beta_0$ on a user-specified scale, such as $\beta_1-\beta_0$ or $\beta_1/\beta_0$, depending on the type of outcome and scientific question of interest. Therefore, we focus on deriving the optimal study design for estimation of $\beta_a$, $a=0,1$.

	We will consider optimal study designs for two treatment mechanisms: observational studies and RCTs. In both settings, we will measure the outcome prospectively, with a slight distinction in whether the treatment of interest, $A$, is readily available in EHR.
	 Specifically, to conduct an observational study, we assume that  $A$ is available in the {\phaseone} EHR sample. For each treatment group $A=a$, we aim to select a study cohort and measure the outcomes $Y(a)$ prospectively.
	To conduct an RCT, we first draw the {\phasetwo} study sample, then randomize the recruited study participants to treatment and control and follow up to measure the outcomes of participants. 
	
	An important observation is that, in both scenarios, 
	$Y(a)$ is observed only for patients who are  selected into the study sample (with $R_2=1$) and assigned with treatment $a$ (with  $A=a$), for $a=0,1$, and $Y(a)$ is missing otherwise.
	Therefore the composite indicator $\mathbb{I}(R_2=1)\mathbb{I}(A=a)$ indicates missingness of the outcome of interest, $Y(a)$.
	For each $a\in\{0,1\}$, let
\begin{align*}
	    &Y^\dag = Y(a)\text{, }~~R_{2a}=\mathbb{I}(R_2=1)\mathbb{I}(A=a)=R_2\mathbb{I}(A=a)
	    \text{, }~~
	    \text{and}\\
	    &\lambda_{2a}(\overline{W})=P(R_{2a}=1\mid \overline{W}, R_1=1),
	\end{align*}
	then there is a correspondence to the basic setting considered in Sections~\ref{sec:prelim}-\ref{sec:RE}, in that $Y^\dag$, $R_{2a}$, and $\lambda_{2a}(\overline{W})$  can be viewed as $Y$, $R_{2}$, and $\lambda_2$, respectively.
	As such, one can derive an estimator for $\beta_a$ that incorporates the
selection mechanism in each phase and the optimal two-phase sampling probability $\lambda_{2a}^*(\overline{W})$ following the  same procedure as Section \ref{subsec:opt}. 
	
	We first  modify the existing assumptions correspondingly.
	\begin{itemize}
	    \item[] 1$^\dag$. Missing at random: $(R_{2},A)\ind Y^\dag \mid R_1,\overline{W}$ and $R_1\ind (Y^\dag, W_1) \mid W_0$;
	    \item[] 2$^\dag$. {\color{black}Positivity: $\epsilon<\lambda_{1}(W_0),\lambda_{2a}(\overline{W})\leq1$ for some $\epsilon > 0$;}
	    \item[] 3$^\dag$. One of the following two sets of models is correctly specified:
	    \begin{enumerate}
	        \item[] (i) Selection probabilities: $\lambda_{1}(W_0)$ and $\lambda_{2a}(\overline{W})$;  
	        \item[] (ii) 
	        Outcome regression model: $E(Y^\dag\mid \overline{W},R_1=1)$ and $f(W_1\mid W_0,R_1=1)$ (thus $E(Y^\dag\mid W_0, R_1=1)$ is also correctly specified).
	    \end{enumerate}
	\end{itemize}
Now replacing $R_2/\lambda_2(\overline{W})$ with $R_{2a}/\lambda_{2a}(\overline{W})$ in Eq. (\ref{eq:influence}) and noticing that $R_{2a}Y^\dag=R_{2a} Y$, 
we have the following estimating function for the mean potential outcome $\beta_a$, $a=0,1$
	\begin{equation*}
	        U_a(O^\dag;\beta_a) = \frac{R_1R_{2a}}{\lambda_1(W_0)\lambda_{2a}(\overline{W})}Y-\frac{R_1R_{2a}-R_1\lambda_{2a}(\overline{W})}{\lambda_1(W_0)\lambda_{2a}(\overline{W})}g(\overline{W})-\frac{R_1-\lambda_1(W_0)}{\lambda_1(W_0)}\tilde{g}(W_0)-\beta_a,
	\end{equation*}
	where $O^\dag=(Y^\dag R_{2a},W_1R_1,W_0)
	$,
	$g(\overline{W})=E(Y^\dag\mid \overline{W},R_1=1)$, and $\tilde{g}(W_0)=E(Y^\dag\mid W_0)$. In fact, $U_a(O^\dag;\beta_a)$ is a special case of Eq. \eqref{eq:generalif}, with $m(Y^\dag,\overline{W};\beta_a)=Y^\dag-\beta_a$.



To derive the optimal design for estimation of $\beta_a$, we consider a general setting where the costs for measuring outcomes for patients exposed to different treatments are allowed to be different. Let $B_a$ denote the total budget for estimating $\beta_a$, $C_{0a}$ denote the initial study cost for estimating $\beta_a$, $C_1$ denote the per-individual cost in {\phaseone} that scales with size of an EHR sample, and $C_{2a}(\overline{W})$ denote the per-individual cost in {\phasetwo} that may depend on auxiliary covariates for patient characteristics. We define {\phaseone} cost of accessing the EHR sample, $n_{ea}C_1$, as follows for an RCT or observational study: in an RCT, $n_{ea}C_1$ is the {\phaseone} cost distributed to the treatment arm $a$ according to {\phasetwo} treatment allocation proportion; in an observational study, $n_{ea}C_1$  is the cost of accessing EHR samples in treatment arm $a$.
By Theorem \ref{thm:fixed_n}, we have the following optimal probability of being selected into the {\phasetwo} sample and being assigned with treatment $a$
\begin{equation*}\label{eq:lambda_thm1ext}
				\lambda^*_{2a}(\overline{W}=\overline{w}) =  \min \left\{1, \frac{B_a-C_{0a}-n_{ea}C_1}{n\lambda_1(w_0)}\frac{\sqrt{{Var(Y^\dag\mid \overline{W},R_1=1)}/{C_{2a}(\overline{w})}}}{E\left\{\sqrt{C_{2a}(\overline{W})Var(Y^\dag\mid \overline{W},R_1=1)}\right\}}\right\}.
\end{equation*} 
Similar to Section \ref{subsec:opt}, $Var(Y^\dag\mid \overline{W},R_1=1)$ can be estimated using a pilot sample following the procedure detailed in Section 2.7 of the Supplementary Material \citep{zhang2022supp}.

Note that
\begin{align}
    \lambda_{2a}(\overline{W}) &= P(R_2=1\mid  A=a, \overline{W},R_1=1)P(A=a \mid \overline{W},R_1=1)\label{eq:lambda2a1}\\
    &=P(A=a \mid R_2=1,\overline{W},R_1=1)P(R_2=1\mid \overline{W},R_1=1)\label{eq:lambda2a2}.
\end{align}
For designing an observational study with treatment $A$ readily available in the {\phaseone} sample, we first estimate $P(A=a \mid \overline{W},R_1=1)$ using the entire EHR data under a pre-specified model, then 
derive the optimal {\phasetwo} sampling probability for each treatment group, i.e., $P(R_2=1\mid  A=a, \overline{W},R_1=1)$, based on Eq. \eqref{eq:lambda2a1}.
For designing an RCT,  because treatment $A$ is randomized, we have $P(A=1 \mid R_2=1,\overline{W}, R_1=1)=c$, where $c\in (0,1)$ is a constant. For a given $c$, we derive the optimal phase-II sampling probability, i.e., $P(R_2=1\mid \overline{W},R_1=1)$ based on Eq. \eqref{eq:lambda2a2}.
We present details on the derivation in Section 2.6 of the Supplementary Material \citep{zhang2022supp}. 

Once the {\phasetwo} sample is drawn and 
$Y(a)$ is measured
, one can estimate $\beta_a$ by solving $n^{-1}\sum_{i=1}^n U_a(O^\dag_i;\beta_a)=0$, with similar estimation procedure as Section \ref{subsec:estbeta}. 
A key step is to estimate $g(\overline{W})$ and $\tilde{g}(W_0)$.
Under Assumption 1$^\dag$ we have $g(\overline{W}) 
= E(Y\mid \overline{W}, R_1=1, A=a)$ and $\tilde{g}(W_0)=E\{g(\overline{W})\mid W_0,R_1=1\}$, hence $g(\overline{W})$ and $\tilde{g}(W_0)$ can be estimated using outcomes measured in  {\phasetwo} sample or from pilot data. We detail the estimation strategy in Section 2.7 of the Supplementary Material, and present the overall algorithm in Section 2.8 of the Supplementary Material \citep{zhang2022supp}. 


We further note that the above estimator of $\beta_a$ is doubly robust in the sense that it is consistent if either the selection probability models  or the outcome regression models are correctly specified, i.e., Assumption 3$^\dag$ holds, which we prove in Section 2.6  of the Supplementary Material \citep{zhang2022supp}.

\subsection{Design and estimation for regression models\label{subsec:reg}}
Another example is the coefficient $\beta$ in a regression model, in which case $m(Y,\overline{W};\beta)=d(\overline{W})\{Y-g(\overline{W};\beta)\}$, where $g(\overline{W};\beta)=E(Y\mid \overline{W};\beta)$ is a user-specified outcome regression model indexed by coefficient $\beta$, and $d(\overline{W})$ is a vector of functions of $\overline{W}$ of the same dimension as $\beta$. Here we briefly discuss extension to the estimation of $\beta$.


Let $\epsilon(\beta)=Y-g(\overline{W};\beta)$. Under Assumptions 1-3, \citet{tsiatis2006semiparametric} proposed an 
augmented inverse probability weighted 
estimator for $\beta$ that solves $E[U(O;\beta)]=0$, where
\begin{equation*} \label{eq:influencereg}
	\begin{split}
	    U(O;\beta)&=
			\frac{R_1R_{2}}{\lambda_{1}(W_0)\lambda_{2}(\overline{W})} d(\overline{W}) \epsilon(\beta)
			-\frac{R_1R_{2}-\lambda_{2}(\overline{W})R_{1}}{\lambda_{1}(W_0)\lambda_{2}(\overline{W})}E\{d(\overline{W}) \epsilon(\beta)\mid \overline{W},R_{1}=1\}\\
			&-\frac{R_{1}-\lambda_{1}(W_0)}{\lambda_{1}(W_0)}E\{d(\overline{W}) \epsilon(\beta)\mid W_0\}.
	\end{split}
	\end{equation*}
The asymptotic variance for the estimator is  given by $V(\lambda_2) = \tau^{-1}(O) Var(U)\{\tau^{-1}(O)\}^\top$, where $\tau(O)=E\{\partial u(O;\beta)/\partial\beta^{\top}\}$ and $u(O;\beta) = R_1R_{2}d(\overline{W})\epsilon(\beta)/\lambda_{1}(W_0)\lambda_{2}(\overline{W})$. Thus one can derive the optimal {\phasetwo} sampling probability $\lambda_{2}^*(\overline{W})$ by minimizing $V(\lambda_2)$ under a budget constraint. 



	\section{Simulation}\label{sec:simulation}
	We conduct simulation studies to assess the performance of the optimal two-phase sampling design and the RR estimator in \textcolor{black}{finite samples.} 
	We first generate a target population of size $n=10,000$. 
	We then emulate the process of selection into the {\phaseone} EHR sample from the target population according to a pre-specified {\phaseone} sampling probability,  
	and the resulting EHR sample size $n_e$ is approximately $5,000$.
	We also randomly draw a pilot sample of size $n_p=200$ from the EHR sample. 
	The data generating mechanism is as follows. 

	\begin{itemize}
		\item $W_0\sim N(0.05, 2)$:  patient characteristics that impact selection into the  {\phaseone}  EHR sample
		\item $R_1\mid W_0 \sim \text{Bernoulli}\{p=\lambda_1(W_0)\}
		$:   indicator for selection into the  {\phaseone}  sample  with 
		\begin{enumerate}
			\item $\lambda_1(W_0) =(1+e^{-W_0})^{-1}$: modest selection bias setting
			\item $\lambda_1(W_0) = 0.9\mathbbm{I}(W_0>0.08)+0.1\mathbbm{I}(W_0\leq0.08)$: extreme selection bias 
		\end{enumerate}
		\item $W_1\sim N(0.05, 2)$: additional patient characteristics observed from the {\phaseone}  sample
		\item $Y\mid \overline{W} \sim N\{E(Y\mid \overline{W};{\boldsymbol \alpha}), Var(Y\mid \overline{W};{\boldsymbol \gamma})\}$: the observed outcome, where $E(Y\mid \overline{W};{\boldsymbol \alpha})=\alpha_0+\alpha_1W_0+\alpha_2W_1$ with ${\boldsymbol \alpha}=(0.1,3,0.01)$, and $Var(Y\mid \overline{W};{\boldsymbol \gamma})=\exp(\gamma_{00}+\gamma_0W_0+\gamma_1W_0^2+\gamma_2W_1+\gamma_3W_1^2)$.  
		We set {$\gamma_{00}=-1.5$, $\gamma_1=0.2$}, $\gamma_2=\gamma_3=0.01$, and we consider three {values of $\gamma_0$: $0.97$, $0.82$, and $-0.64$, which correspond to the scenarios where the PVE is low ($0.2$), moderate ($0.5$), and high ($0.8$), respectively.}
	\end{itemize}
	The model parameters $\boldsymbol{\alpha}$ and $\boldsymbol{\gamma}$ are estimated via maximum likelihood estimation  as detailed in Section 2.4  of the Supplementary Material \citep{zhang2022supp}. 
	We further define the total budget and various costs as follows: the total budget is $B=100,000$, the initial study cost is $C_0=50,450$, the per-individual cost for obtaining the EHR data is $C_1=0.01$, and the auxiliary-specific per-individual cost for measuring the outcome of interest is $C_2(\overline{W})=100$.

	We compare the performance of four approaches  listed in Table~\ref{tab:simu_approach} which differ in sampling design and estimation method.
	
	\begin{table}[!ht]
		\caption{
		Four approaches for sampling design and parameter estimation. 
		\underline{Approach 1:} (naive) measure outcomes from a random sample of \textcolor{black}{the {\phaseone}} sample and estimate $\beta$ using the sample mean. 
		\underline{Approach 2:} (Random sampling and the RR estimator) measure outcomes from a random sample of \textcolor{black}{the {\phaseone}} sample and estimate $\beta$ using the RR estimator. 
		\underline{Approach 3:} (Optimal sampling and the RR estimator)
		model {\phaseone} selection bias $\lambda_1(W_0)$, then measure outcomes based on our proposed optimal two-phase sampling design, and estimate $\beta$ using the RR estimator.}
		\label{tab:simu_approach}
		\begin{center}
		\resizebox{0.99\linewidth}{!}{
    \begin{threeparttable}
			\begin{tabular}{ccccccc}
				\hline 
				\multirow{2}{*}{\textbf{Approach}} & \multirow{2}{*}{\textbf{Legend}} & \multicolumn{2}{c}{\textbf{Method for}} & \multirow{2}{*}{\textbf{Estimation}} & \multicolumn{2}{c}{\textbf{Model Specification for}}\tabularnewline
				\cline{3-4} \cline{4-4} \cline{6-7} \cline{7-7} 
				& & \textbf{Phase-I Selection Bias} & \textbf{Phase-II Sampling Design} &  & \textbf{$E[Y\mid W_{0}]$ and $E[Y\mid\overline{W},R_{1}=1]$} & \multirow{1}{*}{\textbf{$Var(Y\mid\overline{W},R_{1}=1)$}}\tabularnewline
				\hline 
				\textbf{1} & $\circ$ & Modeling$^{\dag}$ & Random & Sample mean & N/A & N/A\tabularnewline
				\textbf{2} & $\vartriangle$ & Modeling & Random & RR estimator & Correct & Correct\tabularnewline
				\textbf{3a} & $+$ & Modeling & Optimal$^{\dag\dag}$ & RR estimator & Correct & Correct\tabularnewline
				\textbf{3b} & $\times$ & Modeling & Optimal & RR estimator & Correct & Misspecified\tabularnewline
				\textbf{3c} & $\diamond$ & Modeling & Optimal & RR estimator & Misspecified & Correct\tabularnewline
				\textbf{3d} & $\triangledown$ & Modeling & Optimal & RR estimator & True model & True model\tabularnewline
				\hline 
		\end{tabular}
		\begin{tablenotes}
        \small
        \item \dag Method 1 for handling selection bias in the {\phaseone}  EHR sample, detailed in Section~\ref{sec:lambda1_method1}. In simulation studies, we use the true $\lambda_1(W_0)$, because  performance of the estimation method has been thoroughly evaluated by \citet{Beesley2019.12.26.19015859} and will be further illustrated in our application study \textcolor{black}{in} Section~\ref{sec:application}. 
		\item\ddag  Method 2 for handeling selection bias in the {\phaseone}  EHR sample, detailed in Section~\ref{sec:lambda1_method2}. 
		\item\dag\dag Approach 3 handles {\phaseone} selection bias by modeling $\lambda_1(W_0)$ with corresponding optimal {\phasetwo} sampling probability presented in Eq. (\ref{eq:lambda_thm1}).
        \end{tablenotes}
	\end{threeparttable}}
	\end{center}
	\end{table}
	We investigate the performance of the above approaches in terms of relative efficiency computed from $50,000$ Monte Carlo replications.
	Relative efficiency is measured by the empirical variance of the estimates from each approach standardized by the empirical variance of the estimates from Approach 2.
	Lower relative efficiency value indicates lower variance of {\color{black}$\widehat{\beta}_{RR}$} and hence a more efficient sampling approach relative to Approach 2.
	
\begin{figure}[!ht]
		\centering
		\includegraphics[width=1\linewidth]{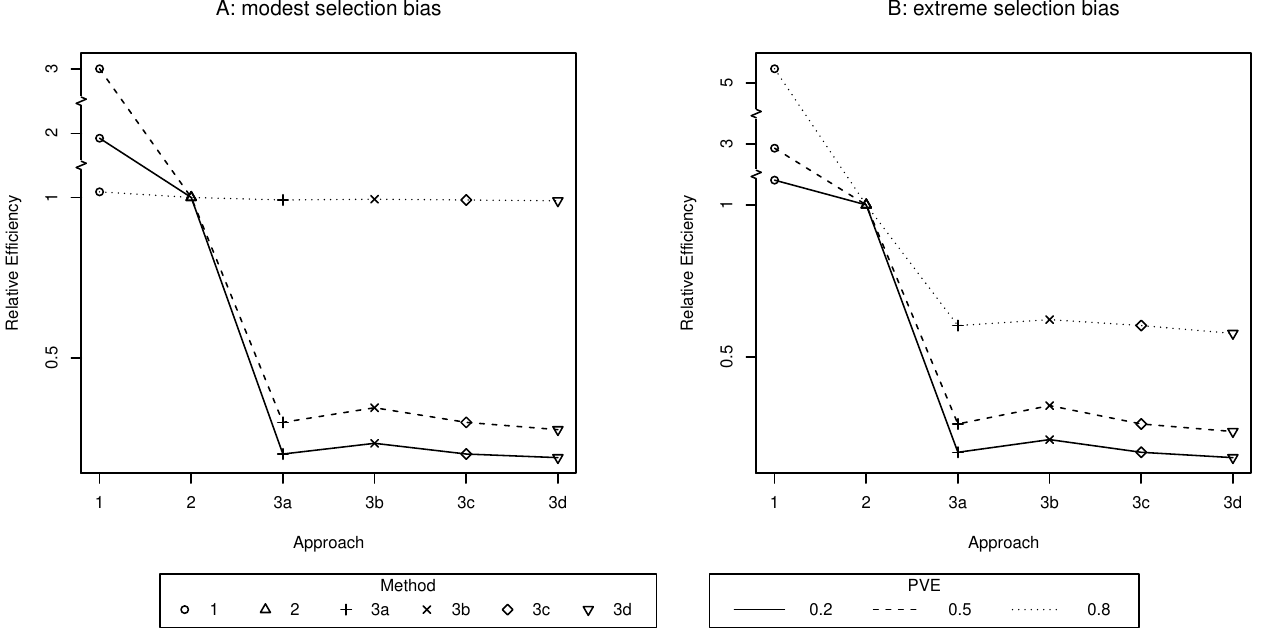}
		\caption{The results of simulation studies. The relative efficiency of Approach 1 and Approach 3a-3c compared to Approach 2, under modest (panel A) and extreme (panel B) selection bias, as well as low ($0.2$), moderate ($0.5$), and high ($0.8$) PVE is presented. The relative efficiency is measured by the Monte Carlo variance of the estimates from each approach divided by that from Approach 2. Lower relative efficiency value indicates smaller estimation variance relative to Approach 2 and hence a more efficient approach. We study the following approaches: (1) naive; (2) random sampling and the RR estimator; (3) optimal sampling and the RR estimator under (3a) correctly specified mean and variance models; (3b) misspecified variance model; (3c) misspecified mean model; and (3d)  true models. PVE stands for the \underline{p}roportion of the \underline{v}ariation in the outcome \underline{e}xplained, which is defined in Section~\ref{sec:RE}.}
		\label{fig:fig_sim}
	\end{figure}

	In Figure \ref{fig:fig_sim}, we present the relative efficiency of Approaches 1-3d compared to Approach 2 under modest and extreme selection bias, as well as low, moderate, and high PVE. 
	Across all PVEs and in both panel\textcolor{black}{s}, we can see that the RE of Approach 1 compared to Approach 2 is larger than one, which indicates that the RR estimator that utilize\textcolor{black}{s} auxiliary information is more efficient than a simple average of the measured outcomes from {\phasetwo}.
	Comparing Approach 3 and Approach 2, the RE is generally smaller than one, which implies that our proposed optimal two-phase sampling design improves efficiency over random sampling. Approach 3b with $Var(Y\mid\overline{W},R_1=1)$ misspecified is the least efficient among all scenarios of Approach 3, which is expected because the optimal sampling design depends on \textcolor{black}{the} estimation of $Var(Y\mid\overline{W},R_1=1)$. Misspecification of the variance model can have \textcolor{black}{a}  notable impact on efficiency, while misspecification of the mean models (Approach 3b) does not have a substantial impact. 
	
	We further investigate the role of PVE which measures the predictive accuracy of the auxiliary covariates \textcolor{black}{$\overline{W}$}. In both panels A and B, the RE of Approach 3 compared to Approach 2 increases with PVE, which is consistent with the conclusion of \textcolor{black}{Corollary}~\ref{cor:total}. We conduct sensitivity analysis under a relatively smaller pilot sample size ($n_p=50$). The results are presented in Figure B4 of  Section 3  of the Supplementary Material \citep{zhang2022supp}. We observe similar efficiency gain from the optimal two-phase sampling approaches particularly under modest selection bias, while the small pilot data may not be sufficient under an extremely biased {\phaseone} sample. We also show the relative efficiency comparing the alternative phase-II sampling design to random sampling under different settings in Section 3 of the Supplementary Material \citep{zhang2022supp}.

	\section{Estimating the prevalence of hypertension among US adults using MGI data\label{sec:application}}
	We estimate the prevalence of hypertension in the US adult population using EHR data from the MGI. A flowchart of the two-phase sampling procedure and data summaries is  presented in Figure~\ref{fig:appdata}. 
	We take the 2017-2018 NHANES data on $6,724$ individuals aged $18$ and older as our external probability sample of the target population to estimate $\lambda_1(W_0)$ following methods in Section~\ref{sec:lambda1_method1}. 
We choose hypertension as our outcome of interest because it is readily measured in MGI and NHANES data, which allows us to validate our results. 
	Using inverse probability of sampling weighting, we estimated from NHANES data that the benchmark prevalence of hypertension in the target population is $39.19\%$.
	
	Let $Y\in\{0,1\}$ denote the indicator of being diagnosed with hypertension, then our parameter of interest $\beta=P(Y=1)$, where the expectation is taken with respect to the distribution of $Y$ in the US adult population. 
	Hypertension status is best determined by chart review, which is time-consuming and resource-intensive. Due to the limitation of resources, we use traditional rule-based phenotyping method instead, which is commonly used  to identify hypertension \citep{chang2016prognostic,fox2014use}.
	We classify hypertension status from MGI data using the following criteria: $Y$ is unknown for patients who did not receive a hypertension diagnostic procedure (defined by the Current Procedural Terminology (CPT) codes listed in Table C2 of   Section 4.1  of the Supplementary Material \citep{zhang2022supp}); for those who received at least one hypertension diagnostic procedure, we define $Y=1$ if there is the presence of at least one hypertension diagnosis code in the patient's record (defined by the International Classification of Diseases (ICD) codes listed in Table C2 of    Section 4.1  of the Supplementary Material \citep{zhang2022supp}), and $Y=0$ otherwise. We take age, sex, and race as $W_0$ and take smoking status and \textcolor{black}{body mass index (BMI)} as $W_1$, which are risk factors associated with hypertension \citep{brown2000body, pinto2007blood}. 
	We categorize BMI as follows: underweight if BMI $<$ 18.5, normal weight if BMI $\in [18.5, 25)$, overweight if BMI $\in [ 25, 30)$, and obese if BMI $\geq 30$. We assume the target population size  is $n=330,000,000$. We consider two scenarios where the total budget $B$ is equal to either $100,000$ or $1,000,000$. We assume  various study costs as follows: the initial study cost is $C_0=10,000$, the  per-individual cost for obtaining the EHR sample is $C_1=0.01$, and the auxiliary-specific per-individual cost for measuring the outcome is $C_2(\overline{W})=100+5\times \text{age}+5\times \text{gender}+5\times \text{race}+5\times \text{BMI}+5\times \text{smoking}$.

	We define the {\phaseone} EHR sample under the following exclusion criteria. 
	We exclude MGI individuals with missing age, gender, race, smoking status, BMI status, and individuals under the age of 18.
	We further exclude individuals who did not receive any diagnostic procedure for hypertension, such that any pilot sample randomly selected from the EHR sample have  outcome  data available for illustration purpose. After data cleaning, the {\phaseone} EHR sample size is $n_e= 80,934$.
	The majority of individuals in the EHR sample are over $60$ ($47.51\%$), female ($53.92\%$), white ($84.49\%$), never-smokers ($54.72\%$), and obese ($42.93\%$).  In addition, hypertension is more prevalent among individuals who are over $60$ ($69.25\%$), male ($54.96\%$), black ($57.58\%$), former smokers ($60.18\%$), and obese ($64.28\%$). As a consequence, the raw prevalence of hypertension is 49.19\% in MGI, which is much higher than the benchmark value 
	and indicates selection bias in the EHR sample.

	We apply Approaches 1-3 investigated in the simulation study. 
	\textcolor{black}{We assume that $Var(Y\mid \overline{W};{\boldsymbol \gamma})=\exp(\gamma_{00}+\gamma_0\text{age}+\gamma_1\text{sex}+\gamma_2\text{race}+\gamma_3\text{BMI}+\gamma_4\text{smoking})$, $E(Y\mid W_0;{\boldsymbol \delta})=\delta_{00}+\delta_0\text{age}+\delta_1\text{sex}+\delta_2\text{race}$, and $E(Y\mid \overline{W};{\boldsymbol \alpha})=\alpha_{00}+\alpha_0\text{age}+\alpha_1\text{sex}+\alpha_2\text{race}+\alpha_3\text{BMI}+\alpha_4\text{smoking}$.}
	We randomly draw $100$ individuals from the EHR sample as the pilot data to estimate parameters in the variance model, \textcolor{black}{while we use the {\phasetwo} study sample to estimate parameters in the mean models.}
	We address selection bias in EHR data by modeling (detailed in Section 3.2.1).
\textcolor{black}{We {choose} to fit a beta regression model which {is} supported by a goodness-of-fit test and the observation that the empirical distribution of the sample probability in the external probability sample {is} unimodal.}
	We obtain estimates of $\beta$ from Approaches 3. 
	We use the sample variance of $50,000$ bootstrapped estimates to calculate the 
	relative efficiencies of Approach 3 versus Approach 1 and Approach 3 versus Approach 2, denoted as $\text{RE}_{\text{3vs1}}$ and $\text{RE}_{\text{3vs2}}$, respectively. 
 \textcolor{black}{We clarify that we aim to apply our optimal sampling to improve efficiency of the RR estimator that accounts for selection bias, and the sampling design itself does not aim to reduce selection bias by creating a study sample with similar characteristics as the target population. Thus, the estimated hypertension prevalence is mainly used to compare the relative efficiency of different sampling approaches rather than to demonstrate reduction of selection bias.} %

\begin{figure}[!h]
		\centering
		\includegraphics[width=\linewidth]{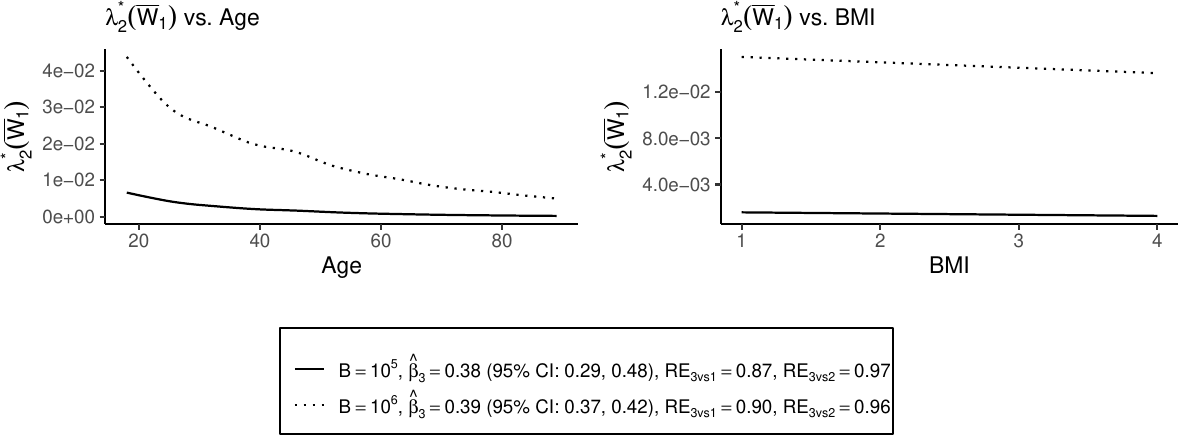}
		\caption{Results of estimating hypertension prevalence in the US adult population using EHR data from the MGI: the relationship between the optimal second phase sampling probability $\lambda_2^*(\overline{W})$ and auxiliary covariates (age and BMI), as well as the relative efficiency comparing different approaches. This figure shows that $\lambda_2(\overline{W})$ tends to decrease with age and BMI and increase with budget. In the legend, $\widehat{\beta}_3$ is the estimate of $\beta$ from Approach 3, $\text{RE}_{\text{3vs1}}$ is the relative efficiency of Approach 3 versus Approach 1, and $\text{RE}_{\text{3vs2}}$ is the relative efficiency of Approach 3 versus Approach 2. 
		}
		\label{fig:fig_app_new}
	\end{figure}

	Figure \ref{fig:fig_app_new} presents smoothed curves for the relationship between the designed $\lambda_2^*(\overline{W})$ from Approach 3 and two selected auxiliary covariates, which are age and BMI. We can see that the optimal sampling probability decreases with both age and BMI, which implies that the proposed method would oversample younger individuals with lower BMI, who are less likely to have hypertension {\citep{brown2000body}}. 
	This is expected because, as discussed previously, the prevalence of hypertension in the EHR sample ($49.19\%$) is higher than that of the target population ($39.19\%$). Our proposed sampling design is able to distinguish the  compositions of patient characteristics between the EHR sample and the target population. 
	As a result, individuals who are less likely to have hypertension are oversampled by our proposed sampling design, which will lead to a final estimate that is lower than the raw prevalence in the EHR sample.
	We also observe that as the total budget $B$ increases, $\lambda_2^*(\overline{W})$ uniformly increases.
	This is also expected because, as budget increases, we can afford a larger study sample and thus individuals in the EHR sample are generally more likely to be sampled.
	
	We also present the estimated hypertension prevalence and relative efficiency comparing different approaches in the legend of Figure \ref{fig:fig_app_new}.
	{\color{black}By Approach 3, when $B=10^5$, the size of the selected study sample is approximately $n_s=100$ and the estimated prevalence is $38.32\%$ ($95\% \text{ CI: }28.90\%, 47.75\%$); when $B=10^6$, the study sample size is approximately $n_s=1,000$ and the estimated prevalence is $39.47\%$ ($95\% \text{ CI: }36.61\%, 42.32\%$).} 
	Compared to the raw estimate from MGI data, these estimates are closer to the benchmark value. The estimated $\text{RE}_{\text{3vs1}}$'s and $\text{RE}_{\text{3vs2}}$'s are all smaller than one, which demonstrates the efficiency gain from both the optimal sampling design and from incorporating auxiliary information in estimation. 
    
	{\color{black}
	We consider the US adult population as the target population in the study above, but one may argue that it is more relevant to generalize to the Michigan adult population. Therefore, we perform a sensitivity analysis (details in   Section 4.2  of the Supplementary Material \citep{zhang2022supp}) to estimate the prevalence of hypertension among Michigan adults to further validate the efficiency gain of the proposed two-phase sampling framework. Compared to the raw estimate from MGI data subject to selection bias, the estimate obtained from Approach 3 is more precise with efficiency gain from our derived optimal sampling design compared to random sampling.
	}

\section{Discussion}\label{sec:discussion}
	
	Electronic health records data open up opportunities for cost-effective patient recruitment \citep{mc2019using}. In this paper, motivated by the observation that recruiting patients from EHR sample constitutes a two-phase sampling framework, we derive the optimal sampling design to minimize the asymptotic variance of an estimator of the mean or mean difference of an outcome of interest that incorporates the selection mechanism in each phase. We extend the two-phase sampling framework proposed by \citet{gilbert2014optimal} to  further account for potential selection bias in  EHR data. 
	We highlight that our proposed method is efficient in design and robust in estimation. First, our proposed two-phase sampling design is more efficient than random sampling, which is shown in asymptotic theory, through finite sample simulation studies, and with an application study to real-world EHR data. Second, the RR estimator is doubly robust in the sense that it is consistent under correct specification of either the {\phaseone} selection mechanism or the outcome models. In addition, by incorporating auxiliary information in all phases, the RR estimator is generally more efficient than a simple weighted average of the outcome. 
 We also extend the proposed method to a general two-stage sampling design framework for estimation of a general estimand, such as the ATE and regression coefficients.
	
	\textcolor{black}{There is a rich literature on optimal two-phase sampling and an evolving literature on patient recruitment using EHR \citep{levis2022double,barrett2019selective,gilbert2014optimal}. 
 \citet{barrett2019selective}  proposed to recruit using EHR such that the distribution of covariate in the study sample is approximately uniformly distributed.  
\cite{levis2022double} considered the scenario where outcomes in the EHR sample are potentially missing not at random and proposed a double sampling design to further collect outcomes. They established identification and derived nonparametric efficient estimators. They ultimately extended the framework to arbitrary coarsening mechanisms, which includes the two-phase sampling framework of \citet{gilbert2014optimal} as a special case.}
It is noteworthy that \citet{gilbert2014optimal}  proposed an optimal three-phase sampling framework in their appendix, which assumed a random {\phaseone} sample and proposed \textcolor{black}{an} optimal sampling design for the second and  third phases.
	Our setting differs from that of \citet{gilbert2014optimal} in that investigators have no control over the selection mechanism of the EHR sample, \textcolor{black}{thus the optimal three-phase sampling framework cannot be directly applied to our setting.}
  \textcolor{black}{We make the following contributions to the literature. First, to our limited knowledge, existing literature did not address the impact of selection bias in EHR on study design. Our work, built upon \cite{gilbert2014optimal}, provides sampling and estimation approaches accounting for selection bias commonly seen in EHR data. Second, we extend the current literature to efficient sampling for the estimation of a general parameter of interest, which covers a wide range of problems such as coefficients in regression models of many kinds, and the average treatment effect in causal inference.
Our proposed method may also be applied to guide the process of selecting a chart review sample to obtain gold standard labels for semi-supervised learning and building phenotyping algorithms \citep{
	 beaulieu2016semi, zhang2022prior}.}
	
	Our proposal has the following limitations. First, EHR systems are designed and optimized for clinical and billing purposes rather than research. As a result, the data collected are subject to a range of issues including missing data, measurement\textcolor{black}{,} and classification error, and confounding, which can undermine the validity of any EHR-based research. 
	Second, in practice, modeling the propensity for an individual to seek care \textcolor{black}{is} likely challenging and may depend on covariates that are not  measured, such as time-varying biomarkers, symptoms, and prior outcomes. The validity of our proposed methods is potentially limited by the amount of prior knowledge on $W_0$.	\textcolor{black}{Third, our positivity assumption for the probability of being selected into the phase-I EHR sample may not be met in practice and should be carefully evaluated. In particular, generalizability may no longer hold when patients with certain characteristic have zero probability of being selected in the EHR sample and such characteristic modifies the outcome model.}
 Lastly, although using pilot data to estimate design parameters is a common approach in two-phase sampling \citep{gilbert2014optimal,mcisaac2015adaptive}, caution should be taken regarding issues such as the potential selection bias within a pilot study, and the need to further account for uncertainty in estimation based on pilot data when making inference on the parameter of interest.
	Methods to address these limitations warrant future research. 
	
	In summary, we believe our proposed sampling and estimation procedure contributes to the two-phase sampling literature and may shed light on data-driven patient recruitment leveraging large-scale EHR data to facilitate biomedical research.
	




\begin{acks}[Acknowledgments]
We thank the Michigan Genomics Initiative participants, Precision Health at the University of Michigan, and the University of Michigan Medical School Data Office for Clinical and Translational Research for providing data storage, management, processing, and distribution services. We thank the Advanced Research Computing Technology Services at the University of Michigan for providing data storage and computing resources. The study protocols were reviewed and determined exempt by the University of Michigan Medical School Institutional Review Board (IRB ID HUM00177982).
\end{acks}



\begin{supplement}
\stitle{Supplement to “Patient recruitment using electronic health records under selection bias: a two-phase sampling framework”}
\sdescription{We provide additional material to support the results in this paper. Section 1 reviews the literature on two-phase sampling design; Section 2.1 provides a derivation of the total expected cost; Section 2.2 includes a proof of the double robustness of the {RR} estimator; Section 2.3 includes a proof of Theorem~\ref{thm:fixed_n}; Section 2.4 details an estimation procedure for the mean outcome under two-phase sampling; Section 2.5 presents a proof of Corollary~\ref{cor:total}; Section 2.6 includes a proof of the double robustness of the {RR} estimator for ATE; Section 2.7 details estimation strategies for the ATE under two-phase sampling; Section 2.8 details the overall algorithm for design and estimation of the ATE; {\color{black}Section 2.9 presents a table summarizing different methods mentioned in this paper.} Section 3 includes additional results of the simulation study; Section 4.1 lists medical codes used for hypertension phenotyping in the application study; Section 4.2 provides a sensitivity analysis for the application study.}
\end{supplement}
\begin{supplement}
\stitle{Source code to “Patient recruitment using electronic health records under selection bias: a two-phase sampling framework”}
\sdescription{We provide R code for implementing the simulation studies and the application study at \url{https://github.com/guanghaozhang/TwoPhaseSampling}}
\end{supplement}


\bibliographystyle{imsart-nameyear} 
\bibliography{main}       

\end{document}


\maketitle
\tableofcontents
\newpage
\section{A Review of Two-Phase Sampling}
Two-phase sampling was first introduced by \cite{neyman1938contribution} to estimate the mean of an outcome that is expensive to measure. 
	\cite{neyman1938contribution} proposed to allocate the {\phaseone} sample into strata based on observed inexpensive auxiliary covariates, sample from each of the strata, and then construct a weighted estimator for the mean outcome.
	\cite{hidiroglou1998use} proposed to incorporate auxiliary covariates obtained at each phase of the two-phase sampling design by leveraging calibrated weights to improve estimation efficiency.
	Similarly, \cite{chatterjee2003pseudoscore} proposed an efficient semiparametric method leveraging  auxiliary variables in the {\phaseone} sample. 
	\cite{gilbert2014optimal} proposed an optimal two-phase sampling design to minimize the asymptotic variance of a semiparametric efficient estimator proposed by \cite{rotnitzky1995semiparametric} that combines weighting and outcome modeling. 
	Overall, the two-phase sampling framework allows one to improve efficiency over random sampling by incorporating auxiliary information. 

	
	The auxiliary information is also extensively used in outcome-dependent sampling (ODS) design. In ODS design, the probability of being selected into the study depends on the outcome of interest \citep{neuhaus1990effect}. 
	For instance, \citet{wang2009causal} proposed methods for causal inference in a two-phase ODS design with stratified phase-II sampling to measure a rich set of covariates.
	\citet{zhou2011semiparametric} proposed a two-phase outcome-auxiliary-dependent sampling design where the sampling probability depends on a continuous outcome.
	\citet{schildcrout2012outcome} proposed an ODS design that incorporates a time-varying auxiliary variable related to the longitudinal binary response. 
	\citet{mcisaac2015adaptive} proposed to leverage internal pilot data to estimate design parameters.

\section{Methods}
\subsection{\textcolor{black}{Derivation} of the expected cost for measuring the outcome in the {\phasetwo} sample and the total expected cost constraint\label{appendix:cost}}
\begin{proof}
	
First, the expected per-person cost for measuring the outcome in the {\phasetwo} sample is 
\begin{align*}
	 & E\{C_2(\overline{W})\mid R_2=1\}\\
	= &\int C_2(\overline{W}) f(\overline{W}\mid R_2=1)d\overline{W}\\
	= &\int C_2(\overline{W}) \frac{f(R_2=1 \mid \overline{W}, R_1=1)f(R_1=1 \mid \overline{W})f(\overline{W})}{f(R_2=1)}d\overline{W}\\
	= &\int \frac{\lambda_{2}(\overline{W})\lambda_{1}(W_0)C_2(\overline{W})f(\overline{W})}{P(R_2=1)}d\overline{W}\\
	= &\frac{n}{n_s}\int \lambda_{2}(\overline{W})\lambda_{1}(W_0)C_2(\overline{W})f(\overline{W})d\overline{W}\\
	= &\frac{n}{n_s}E\{\lambda_{2}(\overline{W})\lambda_{1}(W_0)C_2(\overline{W})\}.
\end{align*}

Then we have an expected cost constraint $B=C_0+n_eC_1+nE\{\lambda_{1}(W_0)\lambda_{2}(\overline{W})C_2(\overline{W})\}$. That is,  in addition to the initial cost $C_0$, the budget will be used to purchase the {\phaseone} electronic health record (EHR) sample ($n_eC_1$) and measure the outcome for individuals in the {\phasetwo} sample recruited from the EHR sample ($nE\{\lambda_{1}(W_0)\lambda_{2}(\overline{W})C_2(\overline{W})\}$).
\end{proof}

\subsection{Double robustness and unbiasedness of the RR estimator}
Recall that we made the following assumptions:
\begin{align}
    &R_1\ind Y\mid W_0\label{assump: o1}\\
    &R_1\ind W_1 \mid W_0\label{assump: o2}\\
    &R_2 \ind Y \mid R_1, \overline{W}\label{assump: o3}
\end{align}

\begin{proof}[Proof of the double robustness and unbiasedness of the RR estimator]
We first rewrite the {estimating equation} for the RR estimator as
\begin{equation*}
    \begin{split}
        U = \frac{R_1R_2-R_1\lambda_2(\overline{W})}{\lambda_1(W_0)\lambda_2(\overline{W})}\left\{Y-E(Y\mid \overline{W},R_1=1)\right\} +\frac{R_1-\lambda_1(W_0)}{\lambda_1(W_0)}\left\{Y-E(Y\mid W_0)\right\}+Y-\beta.
    \end{split}
\end{equation*}

The estimating function with postulated models is
\small
\begin{equation*}
    \begin{split}
        U^* = \frac{R_1R_2-R_1\lambda^*_2(\overline{W})}{\lambda^*_1(W_0)\lambda^*_2(\overline{W})}\left\{Y-E^*(Y\mid \overline{W},R_1=1)\right\} +\frac{R_1-\lambda^*_1(W_0)}{\lambda^*_1(W_0)}\left\{Y-E^*(Y\mid W_0)\right\}+Y-\beta.
    \end{split}
\end{equation*}

\begin{equation*}
    \begin{split}
        E(U^*)=&E\left[\frac{R_1R_2-R_1\lambda^*_2(\overline{W})}{\lambda^*_1(W_0)\lambda^*_2(\overline{W})}\left\{Y-E^*(Y\mid \overline{W},R_1=1)\right\} +\frac{R_1-\lambda^*_1(W_0)}{\lambda^*_1(W_0)}\left\{Y-E^*(Y\mid W_0)\right\}+Y-\beta\right]\\
        =& E\left[\frac{R_1R_2-R_1\lambda^*_2(\overline{W})}{\lambda^*_1(W_0)\lambda^*_2(\overline{W})}\left\{Y-E^*(Y\mid \overline{W},R_1=1)\right\}\right]
        +E\left[\frac{R_1-\lambda^*_1(W_0)}{\lambda^*_1(W_0)}\left\{Y-E^*(Y\mid W_0)\right\}\right]+E\left[Y\right]-\beta\\
        =& E\left[E\left[\frac{R_1R_2-R_1\lambda^*_2(\overline{W})}{\lambda^*_1(W_0)\lambda^*_2(\overline{W})}\left\{Y-E^*(Y\mid \overline{W},R_1=1)\right\}\mid \overline{W}\right]\right] \\
        & + E\left[E\left[\frac{R_1-\lambda^*_1(W_0)}{\lambda^*_1(W_0)}\left\{Y-E^*(Y\mid W_0)\right\}\mid W_0\right]\right] + 0\\
        \stackrel{(\ref{assump: o1})}{=}& E\left[E\left[E\left[\frac{R_1R_2-R_1\lambda^*_2(\overline{W})}{\lambda^*_1(W_0)\lambda^*_2(\overline{W})}\left\{Y-E^*(Y\mid \overline{W},R_1=1)\right\}\mid \overline{W},R_1\right]\mid \overline{W}\right]\right]\\
        & + E\left[E\left\{\frac{R_1-\lambda^*_1(W_0)}{\lambda^*_1(W_0)}\mid W_0\right\}E\left\{Y-E^*(Y\mid W_0)\mid W_0\right\}\right] \\
        =& E\left[E\left[\frac{R_1R_2-R_1\lambda^*_2(\overline{W})}{\lambda^*_1(W_0)\lambda^*_2(\overline{W})}\left\{Y-E^*(Y\mid \overline{W},R_1=1)\right\}\mid \overline{W},R_1=1\right]P(R_1=1\mid \overline{W})\right] + 0\\
        & + E\left[\frac{P(R_1\mid W_0)-\lambda^*_1(W_0)}{\lambda^*_1(W_0)}E\left\{Y-E^*(Y\mid W_0)\mid W_0\right\}\right] \\
        \stackrel{(\ref{assump: o2})(\ref{assump: o3})}{=}& E\left[E\left\{\frac{R_1R_2-R_1\lambda^*_2(\overline{W})}{\lambda^*_1(W_0)\lambda^*_2(\overline{W})}\mid \overline{W},R_1=1\right\}E\left\{Y-E^*(Y\mid \overline{W},R_1=1)\mid \overline{W},R_1=1\right\}P(R_1=1\mid W_0)\right] \\
        & + E\left[\frac{P(R_1\mid W_0)-\lambda^*_1(W_0)}{\lambda^*_1(W_0)}\left\{E\left(Y\mid W_0\right)-E^*(Y\mid W_0)\right\}\right] \\
        =&E\left[\frac{P(R_2=1\mid \overline{W},R_1=1)-\lambda^*_2(\overline{W})}{\lambda^*_1(W_0)\lambda^*_2(\overline{W})}\lambda_1(W_0)\left\{E\left(Y\mid \overline{W},R_1=1\right)-E^*(Y\mid \overline{W},R_1=1)\right\}\right] \\
        & + E\left[\frac{P(R_1\mid W_0)-\lambda^*_1(W_0)}{\lambda^*_1(W_0)}\left\{E\left(Y\mid W_0\right)-E^*(Y\mid W_0)\right\}\right]
    \end{split}
\end{equation*}
\normalsize


By Assumption (1), we have 
$$E(Y\mid \overline{W},R_1=1)=E(Y\mid \overline{W},R_1=1,R_2=1)=E(Y\mid \overline{W},R_2=1).$$

We also have 
\begin{align}
E(Y\mid W_0)\stackrel{Y\ind R_1\mid W_0}{=}&E(Y\mid W_0,R_1=1)\\
{=}&\int_{w_1} E(Y\mid W_0,W_1,R_1=1)f(w_1 \mid W_0,R_1=1)dw_1 \nonumber\\
\stackrel{Y\ind R_2\mid \overline{W},R_1=1}{=}&\int_{w_1} E(Y\mid W_0,W_1,R_1=1, R_2=1)f(w_1 \mid W_0,R_1=1)dw_1 \nonumber\\
\stackrel{R_2=1 \text{ implies } R_1=1}{=}&\int_{w_1} E(Y\mid W_0,W_1,R_2=1)f(w_1 \mid W_0,R_1=1)dw_1.
\end{align}

We conclude that $E(U^*)=0$  if either (1) $\lambda^*_1(W_0)$ and $\lambda_2^*(\overline{W})$ are correctly specified
or (2) $E^*(Y\mid \overline{W},R_2=1)$ and $f(W_1\mid W_0,R_1=1)$ are correctly specified (such that $E^*(Y\mid W_0)$ is also correct). 

\end{proof}
\subsection{Derivation of the optimal {\phasetwo} sampling probability \label{appendix:fixed_n}}

\begin{proof}[Proof of Theorem 1]
We start with a simplified setting for discrete auxiliary variables.
We assume that discrete random variables $W_0$ and $W_1$ take values in a finite set $\Omega_0$ and $\Omega_1$ respectively, and we let $\Omega = \Omega_0 \times \Omega_1$. 
Our objective is to find the optimal design that minimizes $V(\lambda_2)/n$ under the budget constraint and the positivity assumption, where
\[
V(\lambda_2)=Var(Y) + E\left[\left\{\frac{1}{\lambda_1(W_0)}-1\right\}Var(Y\mid W_0)\right] 
+ E\left[\left\{\frac{1}{\lambda_2(\overline{W})}-1\right\}\frac{1}{\lambda_1(W_0)}Var(Y\mid \overline{W},R_1=1)\right].
\]
      
      Specifically, we have the following optimization problem
\begin{tcolorbox}[add to width=0cm]
\begin{align} \label{eq:conv}
\arg\min_{\lambda_2(\overline{w})}\;\;  \frac{V(\lambda_2)}{n}  &=\frac{1}{n}Var\{E(Y\mid W_0)\}  \nonumber\\
&  + \frac{1}{n}\sum_{w_0 \in \Omega_0}\frac{Var(Y\mid W_0=w_0)-E\{Var(Y\mid\overline{W},R_1=1)\mid W_0=w_0\}}{\lambda_1(w_0)}P(W_0=w_0) \\
& + \frac{1}{n}\sum_{\overline{w} \in \Omega}\frac{Var(Y\mid \overline{W}=\overline{w},R_1=1)}{\lambda_1(w_0)\lambda_2(\overline{w})}P(\overline{W}=\overline{w}) \nonumber\\
\text{s.t.}\;\; & 0<\lambda_2(\overline{W})\leq1\text{ for all }\overline{w} \in \Omega \nonumber\\
&  C_0+n_eC_1+n\sum_{\overline{w} \in \Omega}\lambda_1(w_0)\lambda_2(\overline{w})C_2(\overline{w})P(\overline{W}=\overline{w})-B =0. \nonumber
\end{align}
\end{tcolorbox}
To simplify notation, we  rewrite $V(\lambda_2)$ as
\begin{equation*}
\begin{split}
V(\lambda_2) &= 
 Var\{E(Y\mid W_0)\} + \sum_{w_0 \in \Omega_0}\frac{v_1(w_0)}{\eta_1(w_0)}p_0(w_0)+\sum_{\overline{w} \in \Omega}\frac{v_2(\overline{w})}{\eta_2(\overline{w})}p_1(\overline{w}),\\
 \text{where $v_1(w_0)$}&= Var(Y\mid W_0=w_0)-E\{Var(Y\mid\overline{W},R_1=1)\mid W_0=w_0\}\\
 v_2(\overline{w})& = Var(Y\mid \overline{W}=\overline{w},R_1=1)\\
 \eta_1(w_0) &=\lambda_1(w_0)\\
 \eta_2(\overline{w})&=\lambda_1(w_0)\lambda_2(\overline{w})\\
 p_0(w_0)&= P(W_0=w_0)\\
 p_1(\overline{w})&= P(\overline{W}=\overline{w}).
\end{split}
\end{equation*}
Then the problem of minimizing $V(\lambda_2)/n$ can be written as
\begin{align*}
\arg\min_{\eta_2}\;\; &\frac{V(\lambda_2)}{n}=\frac{1}{n}Var\{E(Y\mid W_0)\} + \frac{1}{n}\sum_{w_0 \in \Omega_0}\frac{v_1(w_0)}{\eta_1(w_0)}p_0(w_0)+\frac{1}{n}\sum_{\overline{w} \in \Omega}\frac{v_2(\overline{w})}{\eta_2(\overline{w})}p_1(\overline{w})\\
\text{s.t.}\;\; & \eta_2(\overline{w})\leq\eta_1(w_0)\text{ for all }\overline{w} \in \Omega\\
&  C_0+n_eC_1+n\sum_{\overline{w} \in \Omega}\eta_2(\overline{w})C_2(\overline{w})p_1(\overline{w})-B =0.
\end{align*}

We first show that the optimization problem above is a convex optimization problem. The second derivative of $V(\lambda_2)/n$  is
\begin{equation*}
	\frac{\partial^2\{V(\lambda_2)/n\}}{\partial\eta^2_2}=\frac{2}{n}\sum_{\overline{w} \in \Omega_1}\frac{v_2(\overline{w})}{\{\eta_2(\overline{w})\}^3}p_1(\overline{w}),
\end{equation*}
where $v_2(\overline{w})\geq 0$, $0<p_1(\overline{w}), \eta_2(\overline{w})<1$. Therefore, $\frac{\partial^2\{V(\lambda_2)/n\}}{\partial\eta^2_2}\geq 0$ and thus the optimization problem above is convex.
 It has been shown that for a convex optimization problem, the point that satisfies the \textcolor{black}{Karush–Kuhn–Tucker conditions (KKT)} conditions is the sufficient and necessary solution for optimality \citep{boyd2004convex}.

We proceed to derive the optimal $\lambda_2(\overline{W})$ that satisfies the KKT conditions to \textcolor{black}{minimize} $V(\lambda_2)/n$.
The KKT conditions can be expressed as 
\begin{align}
0 &< \eta_2(\overline{w}), \overline{w} \in \Omega \label{eq:B1}\\
0 &= h(\eta_2) \label{eq:B2}\\
0 &\geq \eta_2(\overline{w})-\eta_1(w_0), w_0 \in \Omega_0, \overline{w} \in \Omega \label{eq:B3}\\
0 &\leq \epsilon(\overline{w}), \overline{w} \in \Omega \label{eq:B4}\\
0 &= \epsilon(\overline{w})f(\eta_2) \label{eq:B5}\\
0 &= \nabla \frac{V(\lambda_2)}{n}+\sum_{\overline{w} \in \Omega}\epsilon(\overline{w})\nabla  f(\eta_2)+\nu\nabla h(\eta_2), \overline{w} \in \Omega   \label{eq:B6}
\end{align}
where  $f(\eta_2)=\eta_2(\overline{w})-\eta_1(w_0)$, $h(\eta_2)= C_0+n_eC_1+nE\{\eta_2(\overline{w})C_2(\overline{w})\}-B$,  and $\nabla f(\eta_2)$ denotes the gradient of a function of $\eta_2$. 

We first focus on conditions (\ref{eq:B4}--\ref{eq:B6}). Because
\begin{align*}
	\nabla \frac{V(\lambda_2)}{n} &= -\frac{1}{n}\sum_{\overline{w} \in \Omega}\frac{v_2(\overline{w})}{\{\eta_2(\overline{w})\}^2}p_1(\overline{w}),\\
	\nabla f(\eta_2) &= 1,\\
	\nabla h(\eta_2) &= n\sum_{\overline{w} \in \Omega}C_2(\overline{w})p_1(\overline{w}),
\end{align*}
 condition (\ref{eq:B6}) is equivalent to
\begin{equation}
\begin{split}
	&-\frac{1}{n}\sum_{\overline{w} \in \Omega}\frac{v_2(\overline{w})}{\{\eta_2(\overline{w})\}^2}p_1(\overline{w}) + \sum_{\overline{w} \in \Omega}\epsilon(\overline{w}) + \nu n\sum_{\overline{w} \in \Omega}C_2(\overline{w})p_1(\overline{w})\\ 
	&=\sum_{\overline{w} \in \Omega}-\frac{v_2(\overline{w})p_1(\overline{w})}{n\{\eta_2(\overline{w})\}^2}+\epsilon(\overline{w}) + \nu n C_2(\overline{w})p_1(\overline{w})=0. \label{eq:B7}
\end{split}
\end{equation}
Set $\epsilon(\overline{w})=0$ for all $\overline{w} \in \Omega$, then by condition (\ref{eq:B7}), we have
\begin{equation}
-\frac{v_2(\overline{w})p_1(\overline{w})}{n\{\eta_2(\overline{w})\}^2}+\nu n C_2(\overline{w})p_1(\overline{w})=0,
\end{equation}
which yields
\begin{equation}
\{\eta_2^*(\overline{w})\}^2=\nu^{-1}\frac{v_2(\overline{w})}{n^{2}C_2(\overline{w})}.\label{eq:B9}
\end{equation}
Substituting (\ref{eq:B9}) into condition (\ref{eq:B2}) we have
\begin{equation}
\nu^{-1/2} = \frac{B-C_0-n_eC_1}{E\left\{\sqrt{C_2(\overline{W})v_2(\overline{W})}\right\}}.\label{eq:B10}
\end{equation}
Then, based on (\ref{eq:B9}) and (\ref{eq:B10}), we have
\begin{equation}
\eta_2^*(\overline{w}) = \frac{1}{n}\sqrt{\frac{v_2(\overline{w})}{C_2(\overline{w})}}\frac{B-C_0-n_eC_1}{E\left\{\sqrt{C_2(\overline{W})v_2(\overline{W})}\right\}}.
\end{equation}
Because $\eta_2(\overline{W})=\lambda_1(W_0)\lambda_2(\overline{W})$, we have
\begin{equation}
\lambda^*_2(\overline{w}) = \frac{1}{n\lambda_1(w_0)}\sqrt{\frac{v_2(\overline{w})}{C_2(\overline{w})}}\frac{B-C_0-n_eC_1}{E\left\{\sqrt{C_2(\overline{W})v_2(\overline{W})}\right\}}. \label{eq:B12}
\end{equation}

We now consider conditions (\ref{eq:B1}--\ref{eq:B3}). Note that by condition (\ref{eq:B2}), we have $nE\{\eta_2(\overline{w})C_2(\overline{w})\}=B-C_0-n_eC_1$. Because per-individual cost is \textcolor{black}{non-negative, i.e., $C_2(\overline{W}) \geq 0$, and $\eta_2(\overline{w})> 0$ by condition (\ref{eq:B1}), we have that $ n_e < (B-C_0)/C_1$}. This is a constraint on the maximal {\phaseone} sample size due to \textcolor{black}{the} limited total budget.
Further, under condition (\ref{eq:B3}), $\lambda^*_2(\overline{w}) \leq 1$.
Therefore, to satisfy this condition, we arrive at 
\begin{equation}
    \lambda^*_2(\overline{w})=\min \left\{1, \frac{1}{n\lambda_1(w_0)}\sqrt{\frac{v_2(\overline{w})}{C_2(\overline{w})}}\frac{B-C_0-n_eC_1}{E\left\{\sqrt{C_2(\overline{W})v_2(\overline{W})}\right\}}\right\}.\label{eq:thm1}
\end{equation}
 In summary, for discrete auxiliary variables, assuming $n_e < (B-C_0)/C_1$, we have that the optimal $\lambda_2(\overline{W})$ that satisfies the KKT conditions is given by (\ref{eq:thm1}).

We then consider the case when the auxiliary variables are continuous. For continuous $W_0$ and $W_1$, we define $f_0(w_0)$ and $f_1(\overline{w})$ as the probability density functions of $W_0$ and $W_1$, respectively. 
Then, the minimization problem is equivalent to 
\begin{align*}
\arg\min_{\eta_2}\;\; &\frac{V(\lambda_2)}{n}=\frac{1}{n}Var\{E(Y\mid W_0)\} + \frac{1}{n}\int\frac{v_1(w_0)}{\eta_1(w_0)}f_0(w_0)dw_0+\frac{1}{n}\int\frac{v_2(\overline{w})}{\eta_2(\overline{w})}f_1(\overline{w})d\overline{w}\\
\text{s.t.}\;\; & 0<\eta_1(w_0)\leq1\\
&\eta_2(\overline{w})\leq\eta_1(w_0)\text{ for all }\overline{w} \in \Omega\\
& h(\eta_2)= C_0+n_eC_1+n\int\eta_2(\overline{w})C_2(\overline{w})f_1(\overline{w})d\overline{w}-B =0.
\end{align*}

In the continuous case, we have the same KKT conditions as the discrete case. Similar to the transition from (\ref{eq:B6}) to (\ref{eq:B7}), we want to interchange the differentiation and integration for 
\begin{align*}
	\nabla \frac{V(\lambda_2)}{n}&=\frac{1}{n}\frac{\partial}{\partial\eta_2}\int\frac{v_2(\overline{w})}{\eta_2(\overline{w})}f_1(\overline{w})d\overline{w}\\
	\text{and }\nabla h(\eta_2)&=n\frac{\partial}{\partial\eta_2}\int\eta_2(\overline{w})C_2(\overline{w})f_1(\overline{w})d\overline{w},
\end{align*}
under the the Leibniz integral rule for differentiation under the integral sign. 
Hence, we assume that $v_2(\overline{w})f_1(\overline{w})/\eta_2(\overline{w})$, $\eta_2(\overline{w})C_2(\overline{w})f_1(\overline{w})$, and their partial derivatives with respect to $\eta_2$ (i.e., $\partial  \{v_2(\overline{w})\linebreak f_1(\overline{w})/\eta_2(\overline{w})\} /\partial \eta_2$ and  $\partial \{\eta_2(\overline{w})C_2(\overline{w})f_1(\overline{w})\}/\partial \eta_2$) are all continuous in $\overline{w}$ and $\eta_2$.

Then, following the same procedure as the discrete case, we can obtain the same optimal {\phasetwo} sampling probability $\lambda^*_2(\overline{w})$ as stated in (\ref{eq:thm1}) under the condition $n_e < (B-C_0)/C_1$.
Therefore, for both discrete and continuous auxiliary variables, assuming $n_e < (B-C_0)/C_1$, we have that the optimal $\lambda_2(\overline{W})$ that satisfies the KKT conditions is given by (\ref{eq:thm1}).
This concludes the proof of Theorem 1.
\end{proof}

\subsection{Estimation methods for parameter of interest \label{appendix:yw0}}
There are three key quantities necessary for optimal sampling and  estimation, which are $E(Y\mid W_0)$, $E(Y\mid \overline{W},R_1=1)$, and $Var(Y\mid \overline{W},R_1=1)$. In this section, we describe strategies to estimate these three quantities.


We first consider the strategy to estimate $Var(Y\mid \overline{W},R_1=1)$, which is a key component of the {\phasetwo} sampling probability $\lambda^*_2(\overline{W})$. Because the pilot sample is a random sample of the {\phaseone} EHR sample with $R_1=1$, we can estimate model parameters in $Var(Y\mid \overline{W},R_1=1)$ using the pilot data. One should check the representativeness of the pilot sample \textcolor{black}{by} comparing marginal distributions (e.g., Kolmogorov–Smirnov test).
Specifically, we specify a parametric variance model such as $Var(Y\mid \overline{W})=\exp(\gamma_{00}+\gamma_0\overline{W}+\gamma_1\overline{W}^2)$ as in our simulation study, and then estimate $(\gamma_{00},\gamma_0,\gamma_1)$ using the residual maximum likelihood (REML) estimation implemented in the \textit{remlscore} function in R package \texttt{statmod} with $Y$ and $\overline{W}$ from the pilot sample \citep{smyth2002efficient}. Then we obtain $\widehat{Var}(Y\mid \overline{W}=\overline{W}_{i}, R_{1i}=1)=\exp(\hat{\gamma}_{00}+\hat{\gamma}_0\overline{W}+\hat{\gamma}_1\overline{W}^2)$. More flexible estimation methods such as the ones proposed in \cite{ruppert1997local} and \cite{fan1998efficient} are also suitable. 

Second, we show that $E(Y\mid \overline{W},R_1=1)$ can be estimated using either the {\phasetwo} data or the pilot data. Because by Assumption (3), $Y\ind R_2\mid \overline{W},R_1=1$, we have  $$E(Y\mid \overline{W},R_1=1)=E(Y\mid \overline{W},R_1=1,R_2=1)=E(Y\mid \overline{W},R_2=1).$$ Therefore, under a prespecified model for $E(Y\mid \overline{W},R_1=1)$, we can estimate model parameters using the {\phasetwo} data. 
Alternatively, because the pilot sample is a random sample of the {\phaseone} EHR sample, we can also estimate $E(Y\mid \overline{W},R_1=1)$ using the pilot data. Therefore,  $E(Y\mid \overline{W},R_1=1)$ can be estimated using either the {\phasetwo} data or the pilot data.

Lastly, we show that $E(Y\mid W_0)$ can be estimated using either the {\phasetwo} data or the pilot data. 
We note that by Assumption (3), 
\begin{align}
E(Y\mid W_0)\stackrel{Y\ind R_1\mid W_0}{=}&E(Y\mid W_0,R_1=1)\label{eq:pilot1}\\
{=}&\int_{w_1} E(Y\mid W_0,W_1,R_1=1)f(w_1 \mid W_0,R_1=1)dw_1 \nonumber\\
\stackrel{Y\ind R_2\mid \overline{W},R_1=1}{=}&\int_{w_1} E(Y\mid W_0,W_1,R_1=1, R_2=1)f(w_1 \mid W_0,R_1=1)dw_1 \nonumber\\
\stackrel{R_2=1 \text{ implies } R_1=1}{=}&\int_{w_1} E(Y\mid W_0,W_1,R_2=1)f(w_1 \mid W_0,R_1=1)dw_1\nonumber\\
=&E\{E(Y\mid \overline{W},R_2=1),W_0,R_1=1\}. \label{eq:phasetwo1}
\end{align}

By (\ref{eq:pilot1}), we can see that $E(Y\mid W_0)$ can be estimated using the pilot data. Alternatively, by (\ref{eq:phasetwo1}), correct specification of both $E(Y\mid W_0,W_1,R_2=1)$ and $f(W_1 \mid W_0,R_1=1)$  implies correct specification of $E(Y\mid W_0)$, and we can estimate $E(Y\mid W_0)$ by estimating $E(Y\mid W_0,W_1,R_2=1)$ using {\phasetwo} data and estimating $f(W_1 \mid W_0,R_1=1)$ using {\phaseone} data. The second approach leverages data with \textcolor{black}{a} larger sample size (and hence may be more efficient) because we often have a larger {\phasetwo}  sample than the pilot sample, and sufficiently large  {\phaseone} EHR data.

We further present our strategy of estimating $E(Y\mid W_0)$ and $E(Y\mid \overline{W},R_1=1)$  under the setting of our simulation study. We first derive the true model for $E(Y\mid W_0)$ based on the data generation model $E(Y\mid \overline{W};{\boldsymbol \alpha})=\alpha_0+\alpha_1W_0+\alpha_2W_1$. Because $W_0$ and $W_1$ are independent in our simulation setting, we have $$E(Y\mid W_0)=E\{E(Y\mid \overline{W};{\boldsymbol \alpha})\mid W_0\}= \alpha_0+\alpha_1W_0+\alpha_2E(W_1)=\alpha'_0+\alpha_1W_0,$$ 
where $\alpha'_0=\alpha_0+\alpha_2E(W_1)$. 
Therefore, under a correctly specified  model $E(Y\mid W_0)=\alpha'_0+\alpha_1W_0$, we estimate $(\alpha'_0,\alpha_1)$ using the robust linear regression MM estimator implemented in the \emph{glmrob} function in R package \texttt{statmod} with $Y$ and $\overline{W}$ from either the {\phasetwo} sample or the pilot sample \citep{yohai1987high,cantoni2001robust,cantoni2004analysis}. Then we obtain $\hat{E}(Y\mid W_{0}=w_{0i})=\hat{\alpha}'_0+\hat{\alpha}_1W_0$.
Similarly, for $E(Y\mid \overline{W},R_1=1)$, under a correctly specified mean model $E(Y\mid \overline{W})=\alpha_0+\alpha_1W_0+\alpha_2W_1$, we estimate $(\alpha_0, \alpha_1,\alpha_2)$ using  $Y$ and $\overline{W}$ from either the {\phasetwo} sample or the pilot sample. Then we obtain $\hat{E}(Y\mid \overline{W}=\overline{W}_{i}, R_{1i}=1)=\hat{\alpha}_0+\hat{\alpha}_1W_0+\hat{\alpha}_2W_1$.

We note that in practice, estimating $\beta$ involves $n^{-1}\sum_{i=1}^{n}\widehat{E}(Y\mid W_{0i})$, which calculates the population mean outcome via imputation. It is in fact unnecessary to have individual-level data on $W_0$ in the target population to compute this term. We propose the following strategies for estimation of the population mean outcome under the two scenarios described in Section 2.1. 
First, if $W_0$ is available for individuals from an external probability sample drawn from the target population with \textcolor{black}{a} known sampling mechanism, then we can estimate this term via inverse probability weighting.
Second, 
if the distribution of $W_0$ in the target population is available from a public data source, then for a categorical variable we can obtain the value of $\widehat{E}(Y\mid W_{0})$ at each category of $W_0$ and then compute a weighted average of group-specific means, whereas for a continuous variable numerical integration can be used.

\subsection{Derivation of the relative efficiency\label{appendix:fixed_n_cor}}
{\color{black}
\begin{proof}[Proof of Corollary 4.1]
\textcolor{black}{We only consider Corollary 4.1 (i), as the proof of Corollary 4.1 (ii) is analogous to that of Corollary 4.1 (i).}

Recall that the relative efficiency (RE) is given by
\begin{equation}
	\text{RE} = \frac{V(\lambda^*_2)}{V(\overline{\lambda}_2)}, \label{eq:B14}
\end{equation}
where $\overline{\lambda}_2 = (B-C_0-n_eC_1)/[n\lambda_1(W_0)E\{C_2(\overline{W})\}]$  is the random {\phasetwo} sampling probability generated from the constraint (\ref{eq:B2}), and 
\[
V(\lambda_2)=Var(Y) + E\left[\left\{\frac{1}{\lambda_1(W_0)}-1\right\}Var(Y\mid W_0)\right] 
+ E\left[\left\{\frac{1}{\lambda_2(\overline{W})}-1\right\}\frac{1}{\lambda_1(W_0)}Var(Y\mid \overline{W},R_1=1)\right].
\]
We have
\scriptsize
\begin{equation*}
\begin{split}
\text{RE}=& \frac{Var(Y) + E^*\{Var(Y\mid W_0)\} 
	+ E\left[\left\{\frac{1}{\lambda^*_2(\overline{W})}-1\right\}\frac{1}{\lambda_1(W_0)}Var(Y\mid \overline{W},R_1=1)\right]
}{Var(Y) + E^*\{Var(Y\mid W_0)\} 
+ E\left[\left\{\frac{1}{\overline{\lambda}_2}-1\right\}\frac{1}{\lambda_1(W_0)}Var(Y\mid \overline{W},R_1=1)\right]
}\\
&= \frac{Var\{E(Y\mid \overline{W},R_1=1)\}+ E\{Var(Y\mid \overline{W},R_1=1)\} + E^*\{Var(Y\mid W_0)\} 
	+ E\left[\left\{\frac{1}{\lambda^*_2(\overline{W})}-1\right\}\frac{1}{\lambda_1(W_0)}Var(Y\mid \overline{W},R_1=1)\right]
}{Var\{E(Y\mid \overline{W},R_1=1)\}+ E\{Var(Y\mid \overline{W},R_1=1)\} + E^*\{Var(Y\mid W_0)\} 
+ E\left[\left\{\frac{1}{\overline{\lambda}_2}-1\right\}\frac{1}{\lambda_1(W_0)}Var(Y\mid \overline{W},R_1=1)\right]
}\\
&=\frac{Var\{E(Y\mid \overline{W},R_1=1)\}+  E^*\{Var(Y\mid W_0)\} 
	+ E\left\{\frac{1}{\lambda^*_2(\overline{W}\lambda_1(W_0)}Var(Y\mid \overline{W},R_1=1)\right\}-E^*\{Var(Y\mid \overline{W},R_1=1)\}
}{Var\{E(Y\mid \overline{W},R_1=1)\}+  E^*\{Var(Y\mid W_0)\} 
+ E\left\{\frac{1}{\overline{\lambda}_2\lambda_1(W_0)}Var(Y\mid \overline{W},R_1=1)\right\}-E^*\{Var(Y\mid \overline{W},R_1=1)\}
}\\
&=\frac{Var\{E(Y\mid \overline{W},R_1=1)\} + E'\{\lambda_1(W_0)\}+  E\left\{\frac{1}{\lambda^*_2(\overline{W})\lambda_1(W_0)}Var(Y\mid \overline{W},R_1=1)\right\}}{Var\{E(Y\mid \overline{W},R_1=1)\} + E'\{\lambda_1(W_0)\}+  E\left\{\frac{1}{\overline{\lambda}_2\lambda_1(W_0)}Var(Y\mid \overline{W},R_1=1)\right\}},
\end{split}
\end{equation*}
\normalsize 
where 
\begin{equation*}
    \begin{split}
        &E^*\{Var(Y\mid W_0)\}  =E\left[\left\{\frac{1}{\lambda_1(W_0)}-1\right\}Var(Y\mid W_0)\right]\\
        &E^*\{Var(Y\mid \overline{W},R_1=1)\}=E\left[\left\{\frac{1}{\lambda_1(W_0)}-1\right\}Var(Y\mid 
        \overline{W},R_1=1)\right]\\
        &E'\{\lambda_1(W_0)\}=E\left[\left\{\frac{1}{\lambda_1(W_0)}-1\right\}\Big\{Var(Y\mid W_0)-Var(Y\mid \overline{W},R_1=1)\Big\}\right].
    \end{split}
\end{equation*}

\textcolor{black}{If $\lambda^*_2=1$, because $\overline{\lambda}_2 \leq 1$, we have $\text{RE} \leq 1$.}

If \begin{equation}
    \lambda^*_2(\overline{w})= \frac{1}{n\lambda_1(w_0)}\sqrt{\frac{v_2(\overline{w})}{C_2(\overline{w})}}\frac{B-C_0-n_eC_1}{E\left\{\sqrt{C_2(\overline{W})v_2(\overline{W})}\right\}},\nonumber
\end{equation}
plugging in $\lambda^*_2$
and $\overline{\lambda}_2$ we have
{\small\begin{align*}
\text{RE}=&\frac{Var\{E(Y\mid \overline{W},R_1=1)\} + E'\{\lambda_1(W_0)\}+  E\left\{\frac{1}{(B-C_0-n_eC_1)\sqrt{\frac{v_2(\overline{w})}{C_2(\overline{w})}}/{\left[nE\{\sqrt{C_2(\overline{W})v_2(\overline{W})}\}\right]  }}Var(Y\mid \overline{W},R_1=1)\right\}}{Var\{E(Y\mid \overline{W},R_1=1)\} + E'\{\lambda_1(W_0)\}+  E\left\{\frac{1}{ (B-C_0-n_eC_1)/\left[nE\{C_2(\overline{W})\}\right]  }Var(Y\mid \overline{W},R_1=1)\right\}}\\
=&\frac{
	Var\{E(Y\mid \overline{W},R_1=1)\} + E'\{\lambda_1(W_0)\}+  \frac{nE\{\sqrt{C_2(\overline{W})v_2(\overline{W})}\}}{B-C_0-n_eC_1} E\left\{
	\sqrt{  \frac{C_2(\overline{w})}{v_2(\overline{w})}   }
	Var(Y\mid \overline{W},R_1=1)\right\}
}
{
	Var\{E(Y\mid \overline{W},R_1=1)\} + E'\{\lambda_1(W_0)\}+  \frac{nE\{C_2(\overline{W})\}}{B-C_0-n_eC_1}E\left\{Var(Y\mid \overline{W},R_1=1)\right\}
}.
\end{align*}}
Recall that by definition $v_2(\overline{w}) = Var(Y\mid \overline{W}=\overline{w},R_1=1)$, we have
\begin{align*}
\text{RE}=&\frac{
	Var\{E(Y\mid \overline{W},R_1=1)\} + E'\{\lambda_1(W_0)\}+  \frac{nE\{\sqrt{C_2(\overline{W})v_2(\overline{W})}\}}{B-C_0-n_eC_1} E\{\sqrt{C_2(\overline{W})v_2(\overline{W})}\}
}
{
	Var\{E(Y\mid \overline{W},R_1=1)\} + E'\{\lambda_1(W_0)\}+  \frac{nE\{C_2(\overline{W})\}}{B-C_0-n_eC_1}E\left\{v_2(\overline{W})\right\}
}\\
=&\frac{
	Var\{E(Y\mid \overline{W},R_1=1)\} + E'\{\lambda_1(W_0)\}+  \frac{n}{B-C_0-n_eC_1} E\left\{\sqrt{C_2(\overline{W})v_2(\overline{W})}\right\}^2
}
{
	Var\{E(Y\mid \overline{W},R_1=1)\} + E'\{\lambda_1(W_0)\}+  \frac{n}{B-C_0-n_eC_1}E\{C_2(\overline{W})\}E\left\{v_2(\overline{W})\right\}
}.
\end{align*}
Recall that the \underline{p}roportion of the \underline{v}ariation in the outcome \underline{e}xplained (PVE) by the covariates $\overline{W}$ in the {\phaseone} EHR sample is defined as $\text{PVE}=Var\{E(Y\mid \overline{W},R_1=1)\}/Var(Y)$. We finally have
\begin{equation}\label{eq:re1}
    \text{RE}=\frac{\text{PVE}*Var(Y) + E'\{\lambda_1(W_0)\} + \frac{n}{B-C_0-n_eC_1}\left[E\left\{\sqrt{C_2(\overline{W})Var(Y\mid \overline{W},R_1=1)}\right\}\right]^2}{\text{PVE}*Var(Y) + E'\{\lambda_1(W_0)\} +\frac{n}{B-C_0-n_eC_1}E\{C_2(\overline{W})\}E\{Var(Y\mid \overline{W},R_1=1)\}},
\end{equation}
where $E'\{\lambda_1(W_0)\}=E\left[(1-\lambda_1(W_0))\Big\{Var(Y\mid W_0)-Var(Y\mid \overline{W},R_1=1)\Big\}/\lambda_1(W_0)\right]$.
According to the Cauchy--Schwarz inequality, if $X$ and $Y$ are random variables and $E(X^2)$ and $E(Y^2)$ exist, then $\left\{E(XY)\right\}^2 \leq E(X^2)E(Y^2)$ \citep{blitzstein2014introduction}. Hence,
\begin{equation}
	\left[E\{\sqrt{C_2(\overline{W})Var(Y\mid \overline{W},R_1=1)}\}\right]^2 \leq E\{C_2(\overline{W})\}E\{Var(Y\mid \overline{W},R_1=1)\}. \label{eq:B15}
\end{equation}
Therefore, we  conclude that $\text{RE} \leq 1$, which means that the optimal two-phase sampling improves efficiency over random sampling.
This concludes the proof of the corollary.

We additionally show that RE increases with PVE. We have
\begin{equation}
\frac{\partial RE}{\partial PVE}=
\frac{\frac{n}{B-C_0-n_eC_1}Var(Y)\left[E\{C_2(\overline{W})\}E\{Var(Y\mid \overline{W},R_1=1)\}-\left[E\{\sqrt{C_2(\overline{W})Var(Y\mid \overline{W},R_1=1)}\}\right]^2\right]}{\left[\text{PVE}*Var(Y) + E'\{\lambda_1(W_0)\} +\frac{n}{B-C_0-n_eC_1}E\{C_2(\overline{W})\}E\{Var(Y\mid \overline{W},R_1=1)\}\right]^2}.
\end{equation}
By (\ref{eq:B15}), $\partial RE/\partial PVE\geq 0$ and
we conclude that RE increases with PVE.

\end{proof}
}
\begin{proof}[Remark for the relative efficiency comparing $\lambda^{{\scriptscriptstyle\text{alt}}}_2$ to $\overline{\lambda}_2$]
{\color{black}
We derive the relative efficiency comparing the alternative {\phasetwo} sampling design with random sampling in {\phasetwo}. Let $R_{1}'\in\{0,1\}$ be the indicator of whether an individual in the target population is selected into the new {\phaseone} sample. Recall that for the alternative two-phase sampling design,
\[
\lambda_1(w_0)=\lambda^{\scriptscriptstyle \text{alt}}_1=n_e'/n,
\]
 and define $v_2'(\overline{w})=Var(Y\mid \overline{W}, R_1'=1)$.
 The corresponding optimal {\phasetwo} sampling probability is 
\begin{equation} \label{eq:thmalt}
\lambda^{{\scriptscriptstyle\text{alt}}}_2=\lambda^{{\scriptscriptstyle\text{alt}}}_2(\overline{w}) = \min \left\{1, \frac{1}{n{\lambda^{\scriptscriptstyle \text{alt}}_1}}\sqrt{\frac{v_2'(\overline{w})}{C_2(\overline{w})}}\frac{B-C_0-n_e'C_1}{E\left\{\sqrt{C_2(\overline{W})v_2'(\overline{W})}\right\}} \right\}.
\end{equation}

According to the definition of relative efficiency,
\begin{equation}
	\text{RE}^{{\scriptscriptstyle\text{alt}}} = \frac{V(\lambda^{{\scriptscriptstyle\text{alt}}}_2)}{V(\overline{\lambda}_2)}, \label{eq:B17}
\end{equation}
where $\overline{\lambda}_2 = (B-C_0-n_eC_1)/[n\lambda_1(W_0)E\{C_2(\overline{W})\}]$  is the random {\phasetwo} sampling probability generated from the constraint (\ref{eq:B2}).

If 
\begin{equation} \label{eq:thmalt1}
\lambda^{{\scriptscriptstyle\text{alt}}}_2= \frac{1}{n{\lambda^{\scriptscriptstyle \text{alt}}_1}}\sqrt{\frac{v_2'(\overline{w})}{C_2(\overline{w})}}\frac{B-C_0-n_e'C_1}{E\left\{\sqrt{C_2(\overline{W})v_2'(\overline{W})}\right\}},\nonumber
\end{equation}
under similar argument as the proof of eq. (\ref{eq:re1}), we have
\begin{equation}
	\text{RE}^{{\scriptscriptstyle\text{alt}}*}=\frac{\text{PVE}*Var(Y) + E'\{\lambda^{\scriptscriptstyle \text{alt}}_1\} + \frac{n}{B-C_0-n_e'C_1}\left[E\left\{\sqrt{C_2(\overline{W})Var(Y\mid \overline{W},R_1'=1)}\right\}\right]^2 }{\text{PVE}*Var(Y) + E'\{\lambda_1(W_0)\} +\frac{n}{B-C_0-n_eC_1}E\{C_2(\overline{W})\}E\{Var(Y\mid \overline{W},R_1=1)\}},
	\end{equation}
	where $E'\{\lambda^{\scriptscriptstyle \text{alt}}_1\}=E\left[(1-\lambda^{\scriptscriptstyle \text{alt}}_1)\Big\{Var(Y\mid W_0)-Var(Y\mid \overline{W},R_1'=1)\Big\}/\lambda^{\scriptscriptstyle \text{alt}}_1\right]$.

We study the relationship between $n_e'$ and $\text{RE}^{{\scriptscriptstyle\text{alt}}*}$. We have
\begin{equation*}
    \frac{\partial \text{RE}^{{\scriptscriptstyle\text{alt}}*}}{\partial n_e'}=-\frac{nA}{{n_e'}^2}+\frac{nBD}{(C-n_e'D)^2},
\end{equation*}
where 
\begin{equation*}
    \begin{split}
        A&=E\left\{Var(Y\mid W_0)-Var(Y\mid \overline{W},R_1'=1)\right\}\\
        B&=\left[E\left\{\sqrt{C_2(\overline{W})Var(Y\mid \overline{W},R_1'=1)}\right\}\right]^2\\
        C&=B-C_0\\
        D&=C_1
    \end{split}
\end{equation*}

When $A\leq 0$, we have ${\partial \text{RE}^{{\scriptscriptstyle\text{alt}}*}}/{\partial n_e'}>0$.

Now we consider the case of $A> 0$. 
When $AD>B$, we have
\begin{equation*}
    \frac{\partial \text{RE}^{{\scriptscriptstyle\text{alt}}*}}{\partial n_e'}<0 \Leftrightarrow n_e'>\frac{ACD+C\sqrt{ABD}}{AD^2-BD} \text{ or }n_e'<\frac{ACD-C\sqrt{ABD}}{AD^2-BD}.
\end{equation*}

When $AD=B$, we have \begin{equation*}
    \frac{\partial \text{RE}^{{\scriptscriptstyle\text{alt}}*}}{\partial n_e'}<0 \Leftrightarrow n_e'<\frac{C}{2D}.
\end{equation*}

When $0<AD<B$, we have
\begin{equation*}
    \frac{\partial \text{RE}^{{\scriptscriptstyle\text{alt}}*}}{\partial n_e'}<0 \Leftrightarrow \frac{ACD-C\sqrt{ABD}}{AD^2-BD}<n_e'<\frac{ACD+C\sqrt{ABD}}{AD^2-BD}.
\end{equation*}
}

\end{proof}

\setcounter{table}{0}
\renewcommand{\thetable}{C\arabic{table}}
\setcounter{figure}{0}
\renewcommand{\thefigure}{B\arabic{figure}}

\subsection{Extension to average treatment effect\label{subsec:ext}}
Suppose the treatment assignment is denoted as $A$, with $A=1$ if a subject received treatment and $0$ otherwise.
Let $\beta_a=E[Y(a)]$ denote the mean potential outcome, let $R_{2a} = \mathbb{I}(R_2=1)\mathbb{I}(A=a)=R_2\mathbb{I}(A=a)$, and let $\lambda_{2a}(\overline{W})=P(R_{2a}=1\mid R_1=1, \overline{W})$ denote the probability of being assigned with treatment $a$ and selected into the phase-II sample from the phase-I sample, $a=0,1$. 
The average treatment effect is a user-defined scale of contrast between $\beta_0$ and $\beta_1$, so we focus on estimating $\beta_a$, $a=0,1$ hereafter.
Recall we made the following identification assumptions: 

\begin{align}
    &(R_2, A) \ind Y(a) \mid R_1,\overline{W}\label{assump:1}\\
    &R_1\ind (Y(a), W_1) \mid W_0\label{assump:2}.
\end{align}
We first examine the double robustness of the proposed estimator based on  the following estimating function 
\begin{equation*}
    \begin{split}
        U_a &= \frac{R_1R_{2a}}{\lambda_1(W_0)\lambda_{2a}(\overline{W})}Y-\frac{R_1R_{2a}-\lambda_{2a}(\overline{W})R_1}{\lambda_1(W_0)\lambda_{2a}(\overline{W})}E(Y\mid \overline{W}, R_{2a}=1)\\
	        &\quad -\frac{R_1-\lambda_1(W_0)}{\lambda_1(W_0)}E\{ E(Y\mid \overline{W}, R_{2a}=1)\mid W_0,R_1=1\}-\beta_a.
    \end{split}
\end{equation*}

\begin{proof}
We first rewrite the estimating function.

Under the consistency assumption, we have
\begin{equation*}
    I(A=a)Y=I(A=a)\left[I(A=1)Y(1) + (1-I(A=1))Y(0)\right]=
    \begin{cases}
      Y(1), & \text{if}\ a=1 \\
      Y(0), & \text{if}\ a=0.
    \end{cases}
  \end{equation*}
That is, 
\begin{equation}
    I(A=a)Y=Y(a).\label{eq:iden}
\end{equation}

With conditional exchangeability (\ref{assump:1}), we have
\begin{equation}\label{eq:condexch1}
\begin{split}
    E(Y\mid A=a,\overline{W}, R_1=1)&\stackrel{(\ref{assump:1})}{=}E(Y(a)\mid \overline{W}, R_1=1)\\
    &\stackrel{(\ref{assump:1})}{=}E(Y(a)\mid \overline{W}, R_1=1, R_{2}=1, A=a)\\
    &=E(Y(a)\mid \overline{W}, R_2=1, A=a)\\
    &=E(Y\mid \overline{W}, R_{2a}=1).
\end{split}
\end{equation}

In addition, we can show that
\begin{equation}\label{eq:condexch0}
    \begin{split}
        E(Y(a) \mid W_0)=&E(Y(a) \mid W_0, R_1=1)\\
        {=}&\int_{w_1} E(Y(a)\mid W_0,W_1,R_1=1)f(w_1 \mid W_0,R_1=1)dw_1 \\
\stackrel{(\ref{assump:1})}{=}&\int_{w_1} E(Y\mid A=a,W_0,W_1,R_1=1)f(w_1 \mid W_0,R_1=1)dw_1\\
=& E\{ E(Y\mid A=a,\overline{W}, R_1=1)\mid W_0,R_1=1\}\\
\stackrel{(\ref{eq:condexch1})}{=}&E\{ E(Y\mid \overline{W}, R_{2a}=1)\mid W_0,R_1=1\}
    \end{split}
\end{equation}

We can then rewrite the estimating function as

\begin{equation*}
    \begin{split}
        U_a\stackrel{(\ref{eq:iden})(\ref{eq:condexch1})(\ref{eq:condexch0})}{=}& \frac{R_1R_2\mathbb{I}(A=a)}{\lambda_1(W_0)\lambda_{2a}(\overline{W})}\left\{Y(a)-E(Y(a)\mid \overline{W}, R_1=1)\right\} \\
        & +\frac{R_1}{\lambda_1(W_0)}\left\{E(Y(a)\mid \overline{W}, R_1=1)-E(Y(a)\mid W_0)\right\}\\
         & + E(Y(a)\mid  W_0)-\beta_a\\
         =& \frac{R_1R_2\mathbb{I}(A=a)-R_1\lambda_{2a}(\overline{W})}{\lambda_1(W_0)\lambda_{2a}(\overline{W})}\left\{Y(a)-E(Y(a)\mid \overline{W}, R_1=1)\right\} \\
         & + \frac{R_1-\lambda_1(W_0)}{\lambda_1(W_0)}\left\{Y(a)-E(Y(a)\mid W_0)\right\} + Y(a) -\beta_a.
    \end{split}
\end{equation*}

The estimating function with postulated models is
\begin{equation*}
    \begin{split}
        U^*_a=&\frac{R_1R_2\mathbb{I}(A=a)-R_1\lambda^*_{2a}(\overline{W})}{\lambda^*_1(W_0)\lambda^*_{2a}(\overline{W})}\left\{Y(a)-E^*(Y(a)\mid \overline{W}, R_1=1)\right\} \\
         & + \frac{R_1-\lambda^*_1(W_0)}{\lambda^*_1(W_0)}\left\{Y(a)-E^*(Y(a)\mid W_0)\right\} + Y(a) -\beta_a.
    \end{split}
\end{equation*}

\small
\begin{equation*}
    \begin{split}
        E(U^*_a)=& E\left[\frac{R_1R_2\mathbb{I}(A=a)-R_1\lambda^*_{2a}(\overline{W})}{\lambda^*_1(W_0)\lambda^*_{2a}(\overline{W})}\left\{Y(a)-E^*(Y(a)\mid \overline{W}, R_1=1)\right\}\right] \\
        & + E\left[\frac{R_1-\lambda^*_1(W_0)}{\lambda^*_1(W_0)}\left\{Y(a)-E^*(Y(a)\mid W_0)\right\}\right] + E\left(Y(a)\right) -\beta_a\\
        =&E\left[E\left[\frac{R_1R_2\mathbb{I}(A=a)-R_1\lambda^*_{2a}(\overline{W})}{\lambda^*_1(W_0)\lambda^*_{2a}(\overline{W})}\left\{Y(a)-E^*(Y(a)\mid \overline{W}, R_1=1)\right\}\mid \overline{W}\right]\right] \\
        & + E\left[E\left[\frac{R_1-\lambda^*_1(W_0)}{\lambda^*_1(W_0)}\left\{Y(a)-E^*(Y(a)\mid W_0)\right\}\mid W_0\right]\right] + 0\\
        \stackrel{(\ref{assump:2})}{=}&E\left[E\left[E\left[\frac{R_1R_2\mathbb{I}(A=a)-R_1\lambda^*_{2a}(\overline{W})}{\lambda^*_1(W_0)\lambda^*_{2a}(\overline{W})}\left\{Y(a)-E^*(Y(a)\mid \overline{W}, R_1=1)\right\}\mid \overline{W}, R_1\right]\mid \overline{W}\right]\right] \\
        & + E\left[E\left\{\frac{R_1-\lambda^*_1(W_0)}{\lambda^*_1(W_0)}\mid W_0\right\} E\left\{Y(a)-E^*(Y(a)\mid W_0)\mid W_0\right\}\right] \\
        =& E\left[E\left[\frac{R_1R_2\mathbb{I}(A=a)-R_1\lambda^*_{2a}(\overline{W})}{\lambda^*_1(W_0)\lambda^*_{2a}(\overline{W})}\left\{Y(a)-E^*(Y(a)\mid \overline{W}, R_1=1)\right\}\mid \overline{W}, R_1=1\right]P(R_1=1\mid \overline{W})\right] + 0\\
        & + E\left[\frac{P(R_1\mid W_0)-\lambda^*_1(W_0)}{\lambda^*_1(W_0)} \left\{E\left(Y(a)\mid W_0\right)-E^*(Y(a)\mid W_0)\right\}\right] \\
        \stackrel{(\ref{assump:1})(\ref{assump:2})}{=}& E\left[E\left\{\frac{R_1R_2\mathbb{I}(A=a)-R_1\lambda^*_{2a}(\overline{W})}{\lambda^*_1(W_0)\lambda^*_{2a}(\overline{W})}\mid \overline{W}, R_1=1\right\}E\left\{Y(a)-E^*(Y(a)\mid \overline{W}, R_1=1)\mid \overline{W}, R_1=1\right\}P(R_1=1\mid W_0)\right]\\
        & + E\left[\frac{P(R_1\mid W_0)-\lambda^*_1(W_0)}{\lambda^*_1(W_0)} \left\{E\left(Y(a)\mid W_0\right)-E^*(Y(a)\mid W_0)\right\}\right] \\
        =&E\left[\frac{P(R_2=1,A=a\mid R_1=1,\overline{W})-\lambda^*_{2a}(\overline{W})}{\lambda^*_1(W_0)\lambda^*_{2a}(\overline{W})}\lambda_1(W_0)\left\{E\left(Y(a)\mid \overline{W}, R_1=1\right)-E^*(Y(a)\mid \overline{W}, R_1=1)\right\}\right]\\
        & + E\left[\frac{P(R_1\mid W_0)-\lambda^*_1(W_0)}{\lambda^*_1(W_0)} \left\{E\left(Y(a)\mid W_0\right)-E^*(Y(a)\mid W_0)\right\}\right].
    \end{split}
\end{equation*}

\normalsize
We conclude that $E(U^*_a)=0$  if either (1) $\lambda^*_1(W_0)$ and $\lambda^*_{2a}(\overline{W})$ are correctly specified or (2) $E^*(Y(a)\mid W_0)$ and $E^*(Y(a)\mid \overline{W}, R_1=1)$ are correctly specified. 

We now present details about deriving the sampling probability.  Similar to Section \ref{appendix:fixed_n} of the Supplementary Material, we  first obtain the asymptotic variance of the RR estimator for $\beta_a$ by $Var(U_a)$ and derive the optimal {\phasetwo} sampling probability $\lambda_{2a}^*$ with KKT conditions:
\begin{equation*}\label{eq:lambda_thm1ext}
				\lambda^*_{2a}(\overline{W}=\overline{w}) =  \min \left\{1, \frac{B_a-C_{0a}-n_{ea}C_1}{n\lambda_1(w_0)}\frac{\sqrt{{Var(Y(a)\mid \overline{W},R_1=1)}/{C_{2a}(\overline{w})}}}{E\left\{\sqrt{C_{2a}(\overline{W})Var(Y(a)\mid \overline{W},R_1=1)}\right\}}\right\},
\end{equation*}
where $n_{ea}C_1$ is defined as follows in an RCT or observational study:
in an RCT, $n_{ea}C_1$ is the {\phaseone} cost of accessing the EHR sample distributed to the treatment arm $a$ according to {\phasetwo} treatment allocation proportion; in an observational study, $n_{ea}C_1$  is the cost of accessing EHR samples in treatment arm $a$.

It should be noted that $Var(Y(a)\mid \overline{W},R_1=1)$ is unknown and needs to be estimated. We detail the estimation strategy for $Var(Y(a)\mid \overline{W},R_1=1)$ in Section \ref{subsec:estext} of the Supplementary Material.

Given the fact that 
\begin{align}
    \lambda_{2a}(\overline{W}) &= P(R_2=1\mid  A=a, \overline{W},R_1=1)P(A=a \mid \overline{W},R_1=1)\label{eq:lambda2a1}\\
    &=P(A=a \mid R_2=1,\overline{W},R_1=1)P(R_2=1\mid \overline{W},R_1=1),\label{eq:lambda2a2}
\end{align}
We consider implementation of the two-phase sampling in two scenarios: observational studies and randomized controlled trials (RCTs). 

To conduct an observational study, we focus on the setting where the treatment $A$ is readily available in the {\phaseone} EHR sample. We first estimate $P(A=a \mid \overline{W},R_1=1)$ using the entire EHR data under a pre-specified model, then for each treatment group $A=a$, we aim to select a study cohort to prospectively measure the outcomes $Y(a)$ based on the optimal {\phasetwo} sampling probability, which is given by $P(R_2=1\mid  A=a, \overline{W},R_1=1)=\lambda_{2a}(\overline{W})/P(A=a \mid \overline{W},R_1=1)$ based on \eqref{eq:lambda2a1}.

To conduct an RCT, we first draw the {\phasetwo} study sample, then randomize the recruited study participants to treatment and control arms and measure their outcomes. By randomization, we know that $P(A=1 \mid R_2=1,\overline{W})$ is a pre-specified constant $c \in (0,1)$. 
Hence, the optimal {\phasetwo} sampling probability is proportional to  $\lambda^*_{2a}(\overline{W})$ and can be given by $P(R_2=1\mid \overline{W},R_1=1)=\lambda^*_{21}(\overline{W})/c=\lambda^*_{20}(\overline{W})/(1-c)$ based on \eqref{eq:lambda2a2}.
Note that the constraint $\lambda^*_{21}(\overline{W})/c=\lambda^*_{20}(\overline{W})/(1-c)$ implies the following:
\small
\begin{equation*}
    \frac{B_1-C_{01}-n_{e1}C_1}{B_0-C_{00}-n_{e0}C_1}\frac{\sqrt{{Var(Y\mid \overline{W},R_2=1,A=1)}/{C_{21}(\overline{w})}}}{E\left\{\sqrt{C_{21}(\overline{W})Var(Y\mid \overline{W},R_2=1,A=1)}\right\}}\frac{E\left\{\sqrt{C_{20}(\overline{W})Var(Y\mid \overline{W},R_2=1,A=0)}\right\}}{\sqrt{{Var(Y\mid \overline{W},R_2=1,A=0)}/{C_{20}(\overline{w})}}}=\frac{c}{1-c},
\end{equation*}
\normalsize
where $n_{ea}C_1$ in an RCT is the {\phaseone} cost of accessing the EHR sample distributed to the treatment arm $a$, $a=0,1$. One way to distribute the cost is by setting $n_{e1}=n_ec$, $n_{e0}=n_e(1-c)$ according to {\phasetwo} treatment allocation proportion.



\end{proof}

\subsection{Estimation methods for parameter of interest in extension to average treatment effect\label{subsec:estext}}
Similar to Section \ref{appendix:yw0} of the Supplementary Material, there are also three key quantities to estimate, which are $E(Y(a)\mid \overline{W}, R_1=1)$, $E(Y(a) \mid W_0)$, and $Var(Y(a)\mid \overline{W},R_1=1)$.

First, we consider the strategy to estimate $Var(Y(a)\mid \overline{W},R_1=1)$, which is necessary to compute the optimal sampling probability $\lambda^*_{2a}$. Because
\begin{equation*}\label{eq:varexp}
\begin{split}
    &Var(Y(a)\mid \overline{W},R_1=1) \\
    =&E\left(Y^2(a)\mid \overline{W}, R_1=1\right)-\left\{E\left(Y(a)\mid \overline{W}, R_1=1\right)\right\}^2\\
    \stackrel{(\ref{assump:1})}{=}&\int y^2f(Y(a)=y\mid \overline{W}, R_1=1)dy-\left\{E\left(Y\mid \overline{W}, R_1=1,A=a\right)\right\}^2\\
    \stackrel{(\ref{assump:1})}{=}&\int y^2f(Y=y\mid \overline{W}, R_1=1,A=a)dy-\left\{E\left(Y\mid \overline{W}, R_1=1,A=a\right)\right\}^2\\
    =&E\left(Y^2\mid \overline{W}, R_1=1,A=a\right)-\left\{E\left(Y\mid \overline{W}, R_1=1,A=a\right)\right\}^2\\
    =&Var(Y\mid \overline{W},R_1=1,A=a),
\end{split}
\end{equation*}
we can estimate this quantity using observed data from a pilot sample, which should be a random sample of the EHR sample that received treatment $a$. In particular, with the pilot sample, one can first specify a parametric variance model and estimate parameters in such a variance model using REML estimation implemented in the \emph{remlscore} function in R package \texttt{statmod}.

Second, we consider estimation strategies for $E(Y(a)\mid \overline{W}, R_1=1)$ and $E(Y(a) \mid W_0)$. The estimation procedure is slightly different between an observational study and an RCT. For an observational study, in Section \ref{subsec:ext} of the Supplementary Material, we have shown that 
\begin{equation*}
    E(Y(a)\mid \overline{W}, R_1=1) =E(Y\mid A=a,\overline{W}, R_1=1).
\end{equation*}
Therefore, $E(Y(a)\mid \overline{W}, R_1=1)$ can be estimated using observed data from a pilot sample, which is a random sample of the EHR sample that received treatment $a$.
For $E(Y(a) \mid W_0)$, in Section \ref{subsec:ext} of the Supplementary Material, we have shown that 
\begin{equation*}
        E(Y(a) \mid W_0)= E\{ E(Y\mid A=a,\overline{W}, R_1=1)\mid W_0,R_1=1\}.
\end{equation*}
Consequently, to estimate $E(Y(a) \mid W_0)$, we can first estimate $E(Y\mid A=a,\overline{W}, R_1=1)$ with a pilot sample, then estimate $f(w_1\mid W_0, R_1=1)$ with the {\phaseone} sample. We can eventually estimate $E(Y(a) \mid W_0)$ by integrating $E(Y\mid A=a,\overline{W}, R_1=1)f(w_1\mid W_0, R_1=1)$ with respect to $w_1$. In practice, one can specify parametric models for $E(Y\mid A=a,\overline{W}, R_1=1)$ and $f(w_1\mid W_0, R_1=1)$, and estimate parameters in these models.

Now we consider estimation strategies for $E(Y(a)\mid \overline{W}, R_1=1)$ and $E(Y(a) \mid W_0)$ in an RCT. In Section \ref{subsec:ext} of the Supplementary Material, we have shown that 
\begin{equation*}
    E(Y(a)\mid \overline{W}, R_1=1)=E(Y\mid \overline{W}, R_{2a}=1).
\end{equation*}
Therefore, $E(Y(a)\mid \overline{W}, R_1=1)$ can be estimated using observed data from the $A=a$ treatment arm of the {\phasetwo} sample.
For $E(Y(a) \mid W_0)$, in Section \ref{subsec:ext} of the Supplementary Material, we have shown that 
\begin{equation*}
        E(Y(a) \mid W_0)=E\{ E(Y\mid \overline{W}, R_{2a}=1)\mid W_0,R_1=1\}.
\end{equation*}
Consequently, to estimate $E(Y(a) \mid W_0)$, we can first estimate $E(Y\mid \overline{W}, R_{2a}=1)$ with the $A=a$ treatment arm of the {\phasetwo} sample, then estimate $f(w_1\mid W_0, R_1=1)$ with the {\phaseone} sample. We can eventually estimate $E(Y(a) \mid W_0)$ by integrating $E(Y\mid \overline{W}, R_{2a}=1)f(w_1\mid W_0, R_1=1)$ with respect to $w_1$. In practice, one can specify parametric models for $E(Y\mid \overline{W}, R_{2a}=1)$ and $f(w_1\mid W_0, R_1=1)$, and estimate parameters in these models.

	\newpage
\subsection{General procedure for two-stage sampling design and estimation for causal inference}
To clarify our proposed method and the difference between design for an observational study and design for  an RCT, we summarize the procedure for two-stage sampling design and estimation for causal inference in the following two figures.
In Figure \ref{fig:algorithmobs}, we present the overall procedure for sampling
design and  estimation for the mean potential outcome in an observational study. In Figure \ref{fig:algorithmrct}, we present the overall procedure for sampling
design and  estimation for the mean potential outcome in an RCT.
\begin{figure}[H]
		\centering
		\tikzstyle{block} = [rectangle, draw, fill=gray!20, 
		text width=6.5 cm, text centered, 
		rounded corners, minimum height=0.75cm]
		\tikzstyle{wideblock} = [rectangle, draw, fill=gray!20, text centered, 
		text width=18 cm, rounded corners, minimum height=0.75cm]
		\tikzstyle{midblock} = [rectangle, draw, fill=gray!20, text centered, 
		text width=8 cm, rounded corners, minimum height=0.75]
		\tikzstyle{line} = [draw, -latex']
		\tikzstyle{vecArrow} = [thick, decoration={markings,mark=at position
			1 with {\arrow[semithick]{triangle 60}}},
		double distance=1.4pt, shorten >= 5.5pt,
		preaction = {decorate},
		postaction = {draw,line width=1.4pt, white,shorten >= 4.5pt}]
		\tikzstyle{innerWhite} = [semithick, black,line width=1.4pt, shorten >= 4.5pt]
		\scalebox{0.8}{\begin{tikzpicture}[node distance = 5cm, auto]
			Place nodes
			\node[wideblock](inp){\textbf{Input}\vspace{-0.4cm} \begin{flushleft}\begin{multicols}{2}$Y(a)$: outcome variable \\ $\overline{W}$: auxiliary covariates \\ $n$: the size of the target population \\ $n_{ea}$: the size of the EHR sample in treatment arm $a$\\ $n_{pa}$: the size of the pilot sample receiving treatment $a$\\ $B_a$: total budget for estimating $\beta_a$\\ $C_{0a}$: the initial study cost for estimating $\beta_a$ \\$C_1$: the constant per-individual cost \\ $C_{2a}(\overline{W})$: per-individual cost depending on auxiliary covariates and treatment \end{multicols}\end{flushleft}};
			\node[wideblock, below=1cm of inp, yshift=-1cm](l2){\textbf{Calculate $\boldsymbol{\lambda^*_{2a}(\overline{W})}$} with ${Var(Y(a)\mid \overline{W}, R_1=1)}$ and $\lambda_1(W_0)$};
			\node[block, left=-6.8cm of l2, xshift=-0.1cm, yshift=1.5cm](var){\textbf{Estimate $\boldsymbol{Var(Y(a)\!\mid\! \overline{W}, \!R_1\!=\!1)}$} using the pilot sample with $Y$ and $A$ available};
			\node[block, right=-6.8cm of l2, xshift=0.1cm, yshift=1.5cm](l1){\textbf{Estimate $\boldsymbol{\lambda_1(W_0)}$} directly or  address selection bias in the EHR sample indirectly};
			\node[wideblock, below=0.5cm of l2, yshift=0cm](eq7){\textbf{Estimate $\boldsymbol{P(A=a\mid \overline{W}, R_1=1)}$} using the entire EHR data};
			\node[wideblock, below=0.5cm of eq7, yshift=0cm](p2){\textbf{Draw the {\phasetwo} sample for the treatment group $a$} according to ${P(R_2=1\mid A=a, \overline{W},R_1=1)}$ that is derived from \eqref{eq:lambda2a1}, then measure $Y$};
			\node[wideblock, below=0.5cm of p2, yshift=0cm](e){\textbf{Estimate $\boldsymbol{E(Y(a)\mid W_0)}$ and $\boldsymbol{E(Y(a)\mid \overline{W}, R_1=1)}$} using the ${A=a}$ treatment arm of the {\phasetwo} sample or the pilot sample};
			\node[wideblock, below=0.5cm of e, yshift=0cm](beta){\textbf{Estimate $\boldsymbol{\beta_a}$} using $\lambda_1(W_0)$, $\lambda^*_{2a}(\overline{W})$, ${E(Y(a)\mid W_0)}$, and ${E(Y(a)\mid \overline{W},R_1=1)}$};
			\node[midblock, below=0.5cm of beta, yshift=0cm](out){\textbf{Output}\\
			$\widehat{\beta_a}$: the estimated mean potential outcome};
			\path [line] (var) -- (l2);
			\path [line] (l1) -- (l2);
			\path [line] (l2) -- (eq7);
			\path [line] (eq7) -- (p2);
			\path [line] (p2) -- (e);
			\path [line] (e) -- (beta);
			
			\draw[vecArrow] (var) -- (l2);
			\draw[vecArrow] (l1) -- (l2);
			\draw[vecArrow] (l2) -- (eq7);
			\draw[vecArrow] (eq7) -- (p2);
			\draw[vecArrow] (p2) -- (e);
			\draw[vecArrow] (e) -- (beta);
			
			\draw[innerWhite] (var) -- (l2);
			\draw[innerWhite] (l1) -- (l2);
			\draw[innerWhite] (l2) -- (eq7);
			\draw[innerWhite] (eq7) -- (p2);
			\draw[innerWhite] (p2) -- (e);
			\draw[innerWhite] (e) -- (beta);
		\end{tikzpicture}}
		
		\caption{General Procedure for Efficient Two-Phase Sampling and Estimation for Observational Studies.\label{fig:algorithmobs}}
	\end{figure}

	\newpage

\begin{figure}[H]
		\centering
		\tikzstyle{block} = [rectangle, draw, fill=gray!20, 
		text width=6.5 cm, text centered, 
		rounded corners, minimum height=0.75cm]
		\tikzstyle{wideblock} = [rectangle, draw, fill=gray!20, text centered, 
		text width=18 cm, rounded corners, minimum height=0.75cm]
		\tikzstyle{midblock} = [rectangle, draw, fill=gray!20, text centered, 
		text width=8 cm, rounded corners, minimum height=0.75]
		\tikzstyle{line} = [draw, -latex']
		\tikzstyle{vecArrow} = [thick, decoration={markings,mark=at position
			1 with {\arrow[semithick]{triangle 60}}},
		double distance=1.4pt, shorten >= 5.5pt,
		preaction = {decorate},
		postaction = {draw,line width=1.4pt, white,shorten >= 4.5pt}]
		\tikzstyle{innerWhite} = [semithick, black,line width=1.4pt, shorten >= 4.5pt]
		\scalebox{0.8}{\begin{tikzpicture}[node distance = 5cm, auto]
			Place nodes
			\node[wideblock](inp){\textbf{Input}\vspace{-0.4cm} \begin{flushleft}\begin{multicols}{2}$Y(a)$: outcome variable \\ $\overline{W}$: auxiliary covariates \\ $n$: the size of the target population \\
						$n_{e}$: the size of the EHR sample \\
						$c$: the pre-specified proportion of subjects randomly assigned to treatment arm $P(A=1 \mid R_2=1,\overline{W})$ \\
						$n_{ea}$: the size of the EHR sample corresponding to arm $a$, which is used to distribute the {\phaseone} cost according to {\phasetwo} treatment allocation proportion, e.g., $n_{e1}=n_ec$, $n_{e0}=n_e(1-c)$\\ $n_{pa}$: the size of the pilot sample receiving treatment $a$\\ $B_a$: total budget for estimating $\beta_a$\\ $C_{0a}$: the initial study cost for estimating $\beta_a$ \\$C_1$: the constant per-individual cost \\ $C_{2a}(\overline{W})$: per-individual cost depending on auxiliary covariates and treatment \end{multicols}\end{flushleft}};
			\node[wideblock, below=1cm of inp, yshift=-1cm](l2){\textbf{Calculate $\boldsymbol{\lambda^*_{2a}(\overline{W})}$} with ${Var(Y(a)\mid \overline{W}, R_1=1)}$ and $\lambda_1(W_0)$};
			\node[block, left=-6.8cm of l2, xshift=-0.1cm, yshift=1.5cm](var){\textbf{Estimate $\boldsymbol{Var(Y(a)\!\mid\! \overline{W}, \!R_1\!=\!1)}$} using the pilot sample with $Y$ and $A$ available};
			\node[block, right=-6.8cm of l2, xshift=0.1cm, yshift=1.5cm](l1){\textbf{Estimate $\boldsymbol{\lambda_1(W_0)}$} directly or  address selection bias in the EHR sample indirectly};
			\node[wideblock, below=0.5cm of l2, yshift=0cm](eq7){\textbf{Estimate $\boldsymbol{P(R_2=1\mid \overline{W}, R_1=1)}$} with $\lambda^*_{2a}(\overline{W})$ and $c$ based on \eqref{eq:lambda2a2}};
			\node[wideblock, below=0.5cm of eq7, yshift=0cm](p2){\textbf{Draw the {\phasetwo} sample} according to ${P(R_2=1\mid \overline{W},R_1=1)}$ and \textbf{randomize the {\phasetwo} sample into treatment arms} according to $c$, and measure $Y$};
			\node[wideblock, below=0.5cm of p2, yshift=0cm](e){\textbf{Estimate $\boldsymbol{E(Y(a)\mid W_0)}$ and $\boldsymbol{E(Y(a)\mid \overline{W}, R_1=1)}$} using the ${A=a}$ treatment arm of the {\phasetwo} sample or the pilot sample};
			\node[wideblock, below=0.5cm of e, yshift=0cm](beta){\textbf{Estimate $\boldsymbol{\beta_a}$} using $\lambda_1(W_0)$, $\lambda^*_{2a}(\overline{W})$, ${E(Y(a)\mid W_0)}$, and ${E(Y(a)\mid \overline{W},R_1=1)}$};
			\node[midblock, below=0.5cm of beta, yshift=0cm](out){\textbf{Output}\\
			$\widehat{\beta_a}$: the estimated mean potential outcome};
			\path [line] (var) -- (l2);
			\path [line] (l1) -- (l2);
			\path [line] (l2) -- (eq7);
			\path [line] (eq7) -- (p2);
			\path [line] (p2) -- (e);
			\path [line] (e) -- (beta);
			
			\draw[vecArrow] (var) -- (l2);
			\draw[vecArrow] (l1) -- (l2);
			\draw[vecArrow] (l2) -- (eq7);
			\draw[vecArrow] (eq7) -- (p2);
			\draw[vecArrow] (p2) -- (e);
			\draw[vecArrow] (e) -- (beta);
			
			\draw[innerWhite] (var) -- (l2);
			\draw[innerWhite] (l1) -- (l2);
			\draw[innerWhite] (l2) -- (eq7);
			\draw[innerWhite] (eq7) -- (p2);
			\draw[innerWhite] (p2) -- (e);
			\draw[innerWhite] (e) -- (beta);
		\end{tikzpicture}}
		
		\caption{General Procedure for Efficient Two-Phase Sampling and Estimation for Randomized Controlled Trials.\label{fig:algorithmrct}}
	\end{figure}

\subsection{Comparison between different methods in the presence of possible selection bias}
{\color{black}In the following Table \ref{tab:mtds}, we compared different methods mentioned in this paper.
Proposed 1 and Proposed 2 are our proposed optimal phase-II sampling, and they differ by how selection bias is handled. Random Sampling 1 and Random Sampling 2 are methods that apply random phase-II sampling, and the difference is that Random Sampling 2 starts from the new phase-I sample after the subsampling approach, which is a random sample of the target population.
\cite{gilbert2014optimal} is to apply Gilbert's optimal sampling design in our EHR setting, ignoring selection bias in the original phase-I EHR sample.
\begin{table}[H]
\resizebox{1\linewidth}{!}{ \begin{threeparttable}
 \begin{flushleft}
\begin{tabular}{cccccc}
\hline
\multirow{2}{*}{Method} &  \multicolumn{2}{c}{Sampling Probability}                    & \multirow{2}{*}{Handling Selection Bias} & \multirow{2}{*}{Estimator}   & \multirow{2}{*}{Relative Efficiency}                                                                                 \\ \cline{2-3}
                         & Phase-I  & Phase-II &  &     &    \\ \hline
Proposed 1     & $\lambda_1(W_0)$  & $\lambda^*_2(\overline{W})$         & Estimation$^\dag$               & Based on Eq. (1)                  & Corollary 4.1 (i)                   \\\hline
Proposed 2 & $\lambda^\text{alt}_1$ & ${\lambda^\text{alt}_2(\overline{W})}^{\dag\dag}$ & Subsampling$^{\dag\dag\dag}$ & Based on Eq. (1) & Corollary 4.1 (ii) \\\hline
Random Sampling 1$^\ddag$ & $\lambda_1(W_0)$  &$\overline{\lambda}_2 $     &     Estimation            & Based on Eq. (1)                & Corollary 4.1 (i)  \\\hline
Random Sampling 2       &$\lambda^\text{alt}_1$         & $\overline{\lambda}$                 & Subsampling & Based on Eq. (1)                  & Corollary 4.1 (ii)    \\\hline
\cite{gilbert2014optimal} & $\lambda_1(W_0)$ & \bigcell{c}{Eq. (8) of\\ \cite{gilbert2014optimal}} & Ignored &  \bigcell{c}{Eq. (12) of \\ \cite{gilbert2014optimal}$^{\ddag\ddag}$} & \bigcell{c}{Eq. (11) of\\ \cite{gilbert2014optimal}$^{\ddag\ddag\ddag}$}\\ \hline
\end{tabular}
\begin{tablenotes}
        \small
        \item \dag Estimation means to directly model phase-I selection mechanism $\lambda_1(W_0)$, then account for selection bias via inverse probability weighting
        \item\dag\dag Now the new phase-I sample is random, phase-II sampling scheme ${\lambda^\text{alt}_2(\overline{W})}$ is analogous to Eq. (8) of \cite{gilbert2014optimal}
        \item\dag\dag\dag Subsampling means to generate a new phase-I sample that is a random sample of the target population, see details in Section 3.2.2
        \item\ddag {\color{black}In the simulation study, Approach 2 corresponds to Random Sampling 1; we additionally consider Approach 1 which ignores selection bias and uses sample mean as the estimator after random phase-II sampling (i.e., $\overline{\lambda}_2$)}
        \item\ddag\ddag \cite{gilbert2014optimal}'s estimator will be biased in the presence of \textcolor{black}{selection bias} in phase-I
        \item\ddag\ddag\ddag Not comparable with other methods
        \end{tablenotes}
\end{flushleft}
\end{threeparttable}}
\caption{Different methods discussed in our paper}\label{tab:mtds}
\end{table}
}

\newpage
\section{Simulation}
{\color{black} For finite sample simulation studies, we present the relative efficiency comparing the alternative phase-II sampling design to random sampling. Specifically, we apply alternative phase-II sampling design and RR estimator together, which is labeled as Approach 4 in the following figures. The following figure presents the results of simulation studies under $n_p=200$.}

 \begin{figure}[H]
	\centering
	\includegraphics[height=8cm, width=16cm]{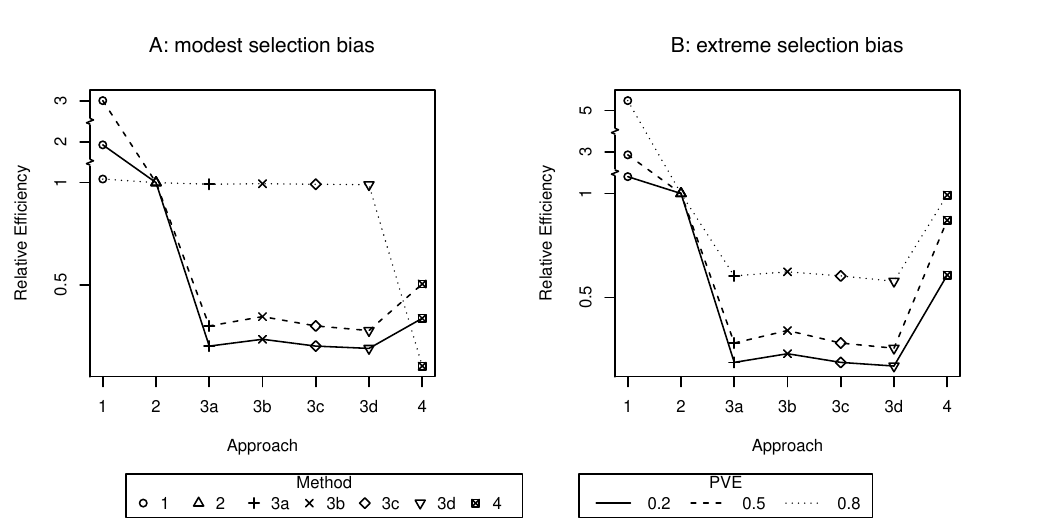}
	\caption{The relative efficiency of Approach 1, Approach 3(a)-(c), and Approach 4 compared to Approach 2, under modest (panel A) and extreme (panel B) selection bias, as well as low ($0.2$), moderate ($0.5$), and high ($0.8$) PVE, and $n_p=200$. The relative efficiency is measured by the Monte Carlo variance of the estimates from each approach divided by that from Approach 2. Lower relative efficiency value indicates smaller estimation variance relative to Approach 2 and hence a more efficient approach. We study the following approaches: (1) naive; (2) random sampling and the RR estimator; (3) optimal sampling and the RR estimator under (3a) correctly specified mean and variance models; (3b) misspecified variance model; (3c) misspecified mean model; (3d)  true models; and (4) alternative optimal sampling and the RR estimator. PVE stands for the \underline{p}roportion of the \underline{v}ariation in the outcome \underline{e}xplained, which is defined in Section 4.}
	\label{fig:fig_sim1}
\end{figure}

For finite sample simulation studies, we conduct sensitivity analysis under a relatively smaller pilot sample size ($n_p=50$). We observe similar efficiency gain from the optimal two-phase sampling approaches particularly under modest selection bias, while the small pilot data may not be sufficient under an extremely biased EHR sample.

\begin{figure}[H]
	\centering
	\includegraphics[height=8cm, width=16cm]{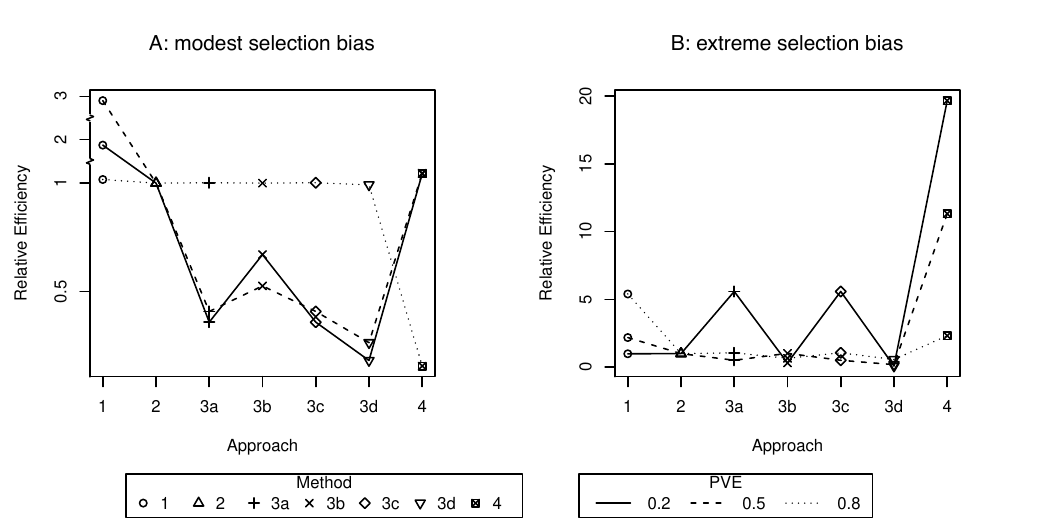}
	\caption{The relative efficiency of Approach 1, Approach 3(a)-(c), and Approach 4 compared to Approach 2, under modest (panel A) and extreme (panel B) selection bias, as well as low ($0.2$), moderate ($0.5$), and high ($0.8$) PVE, and $n_p=200$. The relative efficiency is measured by the Monte Carlo variance of the estimates from each approach divided by that from Approach 2. Lower relative efficiency value indicates smaller estimation variance relative to Approach 2 and hence a more efficient approach. We study the following approaches: (1) naive; (2) random sampling and the RR estimator; (3) optimal sampling and the RR estimator under (3a) correctly specified mean and variance models; (3b) misspecified variance model; (3c) misspecified mean model; (3d)  true models; and (4) alternative optimal sampling and the RR estimator. PVE stands for the \underline{p}roportion of the \underline{v}ariation in the outcome \underline{e}xplained, which is defined in Section 4.}
	\label{fig:fig_sim2}
\end{figure}


\section{Application}\label{appendix:code}

\subsection{Determination of hypertension status in MGI data}
We use CPT-4 codes to define hypertension diagnostic procedures, and we use International Classification of Diseases (ICD) codes (ICD-9/10) codes to define whether the MGI individual was diagnosed with hypertension. With CPT-4 and ICD-9/10 codes, we can follow the process detailed in Section 7 to determine whether the MGI individual's hypertension status is diagnosed, undiagnosed, or missing.

\begin{table}[H]
\centering
\begin{tabularx}{\textwidth}{XX}
\hline
CPT-4 Codes & ICD-9/10 Codes \\
\hline 
93784, 93786, 93788, 93790, 99473, 99474, 90791, 90792, 90832, 90834, 90837, 90839, 90845, 90880, 92002, 92004, 92012, 92014, 96118, 99201, 99202, 99203, 99204, 99205, 99212, 99213, 99214, 99281, 99282, 99283, 99284, 99285, 99215, 99304, 99305, 99306, 99307, 99308, 99309, 99310, 99318, 99324, 99325, 99326, 99327, 99328, 99334, 99335, 99336, 99337, 99340, 99341, 99342, 99343, 99344, 99345, 99347, 99348, 99349, 99350, D7140, D7210, G0101, G0402, G0438, G0439 & 401.0, 401.1, 401.9, 402.00, 402.01, 402.10, 402.11, 402.90, 402.91, 403.00 , 403.01, 403.10, 403.11, 403.90, 403.91, 404.00, 404.01, 404.02, 404.03, 404.10, 404.11, 404.12, 404.13, 404.90, 404.91, 404.92, 404.93, 405.01, 405.09, 405.11, 405.19, 405.91, 405.99, I10, I11.0, I11.9, I12.0, I12.9, I13.0, I13.10, I13.11, I13.2, I15.0, I15.1, I15.2, I15.8, I15.9\\
\hline 
\end{tabularx}
\caption{CPT-4 Codes and ICD-9/10 Codes Used to Determine MGI Individual Hypertension Status\label{tab:appcode}}
\end{table}

\subsection{Sensitivity analysis}
	In the sensitivity analysis, we estimate the prevalence of hypertension among Michigan adults (age $\geq$ 18) to further validate the efficiency gain of the proposed two-phase sampling framework.
	
	We first construct an external probability sample of the Michigan adult population using the joint distribution of age, sex, and race provided by National Center for Health Statistics (NCHS) of Centers for Disease Control and Prevention (CDC) \citep{syoa}. Unfortunately, hypertension status is not available for such an artificial external sample and we cannot obtain a benchmark prevalence of hypertension in the target population. The Behavioral Risk Factor Surveillance System of CDC estimated that 34.70\% of Michigan adults have been told that they have high blood pressure in 2017, and the number increased to 35.10\% in 2019 \citep{brfss}. We then take the simple average of the two estimates (34.90\%) as the estimated hypertension prevalence among Michigan adults in 2018. Note that this estimated prevalence is not a benchmark for this sensitivity study and can only be considered as a reference.
	
	Almost all of the settings remain the same as the application study in Section 7, except that the target population size is now $n=7,831,247$, which is the size of Michigan adult population based on NCHS data. We also apply Approaches 1-4 in the simulation study to estimate $\beta$ and use the sample variance of 50,000 bootstrapped estimates to calculate relative efficiencies.
	With $B=10^5$, by Approach 3, the recruited study sample size is $n_s\approx 180$, the estimated prevalence is $40.48\%$ ($95\% \text{ CI: }32.83\%, 48.13\%$) with $\text{RE}_{\text{3vs1}}=0.85$ and $\text{RE}_{\text{3vs2}}=0.97$.
	Similar to the application study in Section 7, the estimate is closer to the reference prevalence than the raw estimate, and we can obtain efficiency gain from both the optimal sampling design and the RR estimator.




\newpage

\bibliographystyle{abbrvnat} 
\bibliography{sup.bib}